\documentclass[useAMS,usenatbib]{mn2e}
\usepackage{graphicx}
\usepackage{color}
\usepackage{multicol}
\usepackage{longtable} 
\usepackage{color, soul}
\title[\emph{Kepler} rotation hare-and-hounds]{Testing the recovery of stellar rotation signals from \emph{Kepler} light curves using a blind hare-and-hounds exercise}

\author[S.~Aigrain et al.]{S.~Aigrain,$^{1}$\thanks{suzanne.aigrain@astro.ox.ac.uk}, J.~Llama,$^{2}$\thanks{joe.llama@lowell.edu} T.~Ceillier,$^{3}$ M.~L.~das Chagas,$^{4}$ J.~R.~A.~Davenport$^{5}$, \newauthor R.~A.~Garc\'ia,$^{3}$ K.~L.~Hay,$^{6,7}$ A.~F.~Lanza,$^{8}$ A.~McQuillan,$^{9}$  T.~Mazeh,$^{9}$ \newauthor J.~R.~de Medeiros$^{4}$ M.~B.~Nielsen,$^{10}$ and T.~Reinhold$^{10}$ \\
$^{1}$Department of Physics,University of Oxford, Oxford, OX1 3RH, UK\\
$^{2}$Lowell Observatory, 1400 West Mars Hill Rd, Flagstaff, Arizona, 86001, USA \\
$^{3}$CEA/DSM, 91191 Gif-sur-Yvette, France \\
$^{4}$Departamento de F{\'i}sica Te{\'o}rica e Experimental,
Universidade Federal do Rio Grande do Norte, 59078-970, Natal,
Brazil\\
$^{5}$Department of Astronomy, University of Washington, Seattle, USA\\
$^{6}$Blackett Laboratory, Imperial College London, London SW7 2AZ, UK\\
$^{7}$SUPA, School of Physics \& Astronomy, North Haugh, St Andrews, Fife, KY16 9SS, UK\\
$^{8}$INAF / Catania Observatory, 95123 Catania, Italy \\
$^{9}$School of Physics \&
Astronomy, Tel Aviv University, 69978 Tel Aviv, Israel \\ 
$^{10}$Institut f{\"u}r Astrophysik, Universit{\"a}t G{\"o}ttingen,
37077 G{\"o}ttingen, Germany}

\begin{document}

\date{Accepted, 13 April 2015. Received, 26 March 2015; in original form, 9 November 2014}

\pagerange{\pageref{firstpage}--\pageref{lastpage}} \pubyear{2014}

\maketitle

\label{firstpage}

\begin{abstract}
  We present the results of a blind exercise to test the recoverability of stellar rotation and differential rotation in \emph{Kepler} light curves. The simulated light curves lasted 1000 days and included activity cycles, Sun-like butterfly patterns, differential rotation and spot evolution. The range of rotation periods, activity levels and spot lifetime were chosen to be representative of the \emph{Kepler} data of solar like stars. Of the 1000 simulated light curves, 770 were injected into actual quiescent \emph{Kepler} light curves to simulate \emph{Kepler} noise. The test also included five 1000-day segments of the Sun's total irradiance variations at different points in the Sun's activity cycle.

  Five teams took part in the blind exercise, plus two teams who participated after the content of the light curves had been released. The methods used included Lomb-Scargle periodograms and variants thereof, auto-correlation function, and wavelet-based analyses, plus spot modelling to search for differential rotation. The results show that the `overall' period is well recovered for stars exhibiting low and moderate activity levels. Most teams reported values within 10\% of the true value in 70\% of the cases. There was, however, little correlation between the reported and simulated values of the differential rotation shear, suggesting that differential rotation studies based on full-disk light curves alone need to be treated with caution, at least for solar-type stars.

The simulated light curves and associated parameters are available online for the community to test their own methods.
\end{abstract}

\begin{keywords}
methods: data analysis -- techniques: photometry -- surveys: Kepler --
stars: rotation -- starspots
\end{keywords}

\section{Introduction}

The rotational modulation of magnetically active regions on the surface of stars produces quasi-periodic variations in their disk-integrated apparent brightness, which have been used for decades to measure rotation periods for young, active stars. The exquisite photometric quality and baseline of space-based telescopes such as \emph{Kepler}, \emph{CoRoT} and \emph{MOST} have made it possible to do this for tens of thousands of moderately active field stars, many of which display sub-millimagnitude variations that would have been undetectable from the ground. The resulting, extensive rotation period catalogs represent an exciting opportunity to test and refine our understanding of stellar angular momentum evolution, and to develop efficient methods for estimating the ages of field stars based on their rotation rate (gyro-chronology). Furthermore, for active stars, the presence of multiple and/or evolving periodicities in the light curves can be used to study phenomena such as differential rotation and active region evolution, which are important tracers of the dynamo mechanisms driving magnetic field generation in those stars.

To make full use of this new information, however, requires a detailed understanding of the reliability and completeness of measurements of mean and differential rotation rates derived from these full-disk space-based light curves. This paper reports on our attempts to address that problem through a hare and hounds exercise, where simulated star-spot signals were injected into \emph{Kepler} light curves of otherwise quiet stars, and several teams independently attempted to measure rotation and differential rotation rate.

\subsection{Field star rotation studies in the \emph{Kepler} era}

While ground-based time-domain photometric surveys have yielded numerous rotation period measurements for stars in young clusters \citep[see e.g.][and citations therein]{ib09,bou+13}, the limited precision and time-sampling achievable from the ground are typically insufficient to detect rotational modulation in older, less active, slowly rotating field stars.  For many years, the Mount Wilson program \citep{wil78,vau+81,bal+96}, which monitored emission in the cores of the Ca {\sc ii} H \& K lines, for large numbers of Sun-like stars over a 20 year period, with a typical time sampling of one point every few days for each star, was the main source of rotation periods measurements for field stars.

This changed with the advent of space-based photometric monitoring platforms such as \emph{MOST}, \emph{CoRoT} and \emph{Kepler}, which have collectively gathered sub-mmag precision photometry over periods ranging from weeks to years, for over two hundred thousand stars. These, and particularly the \emph{Kepler} light curves, which have the longest baseline and highest precision, have enabled a number of large-scale rotation period studies for field main-sequence stars: \citet{bas+11,aff+12,nie+13,mcq+13a,mcq+13b,mcq+14,rei+13,gar+14}. These studies have produced intriguing results, such as the existence of a bimodality in the period distribution for K and M stars, the sharp upper envelope of the period-mass relation, and the existence of a small but significant population of rapid rotators ($P<10$ days). Some of these studies relied on the Lomb-Scargle periodogram, which has been widely used for rotation period measurements for decades, while others have introduced new methods based for example on the autocorrelation function or a wavelet transform of the light curves.

One of the goals of the present paper was to test the sensitivity and
reliability of these different methodologies on realistic but
simulated \emph{Kepler} data, where the ground truth is known.

\subsection{Differential rotation measurements}

Going beyond measurements of the mean or overall rotation periods, a number of studies have also looked for, and reported signs of, surface differential rotation in long-baseline, disk integrated stellar light curves. The presence of multiple, close but distinct peaks in the Fourier transform or periodogram, and the associated `beat' patterns, which are often observed in the light curves of active stars, are often interpreted as signposts of differential rotation. 

Efforts to measure differential rotation from light curves can be divided into two broad classes. One class uses the broadening or splitting of peaks in the Fourier transform or periodogram of the chromospheric activity indicator measurements or light curve \citep[see e.g.][]{bal+85,don+92,rei+13}. The other class relies on fitting a spot model to the light curve, where the spots may have different rotation periods \citep[see e.g.][]{str+92,cro+06,lan+09,gon+10,lan+14a}.  Once again, these differential rotation studies have yielded intriguing results, such as a trend for increased rotational shear ($\delta P/P$) for increased stellar effective temperature. 

The two classes of methods used have different advantages and disadvantages. Fourier-domain methods can be applied systematically for large numbers of stars, but the relationship between the measured peak width or separation and physical quantities of interest such as the rotational shear is non-trivial. Spot-modelling is computationally intensive, and therefore restricted to smaller samples, but in principle it has a more direct physical interpretation. In practice, however, spot-modelling is highly degenerate, even when using only a few spots. This was recently demonstrated by \citet{wal+13}, who identified degeneracies between stellar inclination and spot latitudes, and showed that differential rotation may be missed if the light curve is dominated by one large active region.

Perhaps even more importantly, both types of approach suffer from a further limitation, which was not explored by \citet{wal+13}: differential rotation is very difficult to distinguish from spot or active region evolution, which can induce very similar beat patterns and broadened or split peaks in the periodogram, and at the very least is an additional source of noise when identifying and modelling individual peaks in the periodogram, or individual star spots \citep{lan+94}.  The second goal of the present study was to test the reliability of differential rotation results reported in the recent literature for stars exhibiting low and moderate levels of activity, again by using realistic simulations including both differential rotation and spot evolution.  

\begin{table*}
  \centering
  \caption{Range and distributions used for the simulation parameters.}
  \label{tab:pardist}
  \begin{tabular}{lll}
    \hline
   Parameter & Range & Distribution \\
\hline \hline
   Activity level $A$ & $0.3$ -- $3 \times$~solar & log uniform \\  
   Activity cycle length $C_{\rm len}$ & $1$ -- $10$~years & log uniform \\  
   Activity cycle overlap $C_{\rm over}$ & $0.1$ -- $3$~years & log
   uniform \\  
   Minimum spot latitude $\theta_{\rm min}$ & $0$ -- $40^{\circ}$ & uniform \\
   Maximum spot latitude $\theta_{\rm max}$ & $\theta_{\rm
     min}$ -- $80^{\circ}$ & uniform in $(\theta_{\rm max}-\theta_{\rm
     min})^{0.3}$ \\
   Inclination $i$ & $0$ -- $90^{\circ}$ & uniform in $\sin^2 i$ \\
   Equatorial rotation period $P_{\rm eq}$ & $10$ -- $50$~days (90\%) &
   log uniform \\ 
   & $1$ -- $10$~days (10\%) & log uniform \\ 
   Rotational shear $\delta\Omega/\Omega_{\rm eq}$ & $0.1$ -- $1$ ($2/3$) & log
   uniform \\
   & $0$ ($1/3$) & \\ 
   Decay timescale $\tau$ & $(1$ -- $10) \times P_{\rm eq}$ &
   log uniform \\ \hline
  \end{tabular}
\end{table*}

\begin{figure*} 
  \centering
  \includegraphics[width=\linewidth]{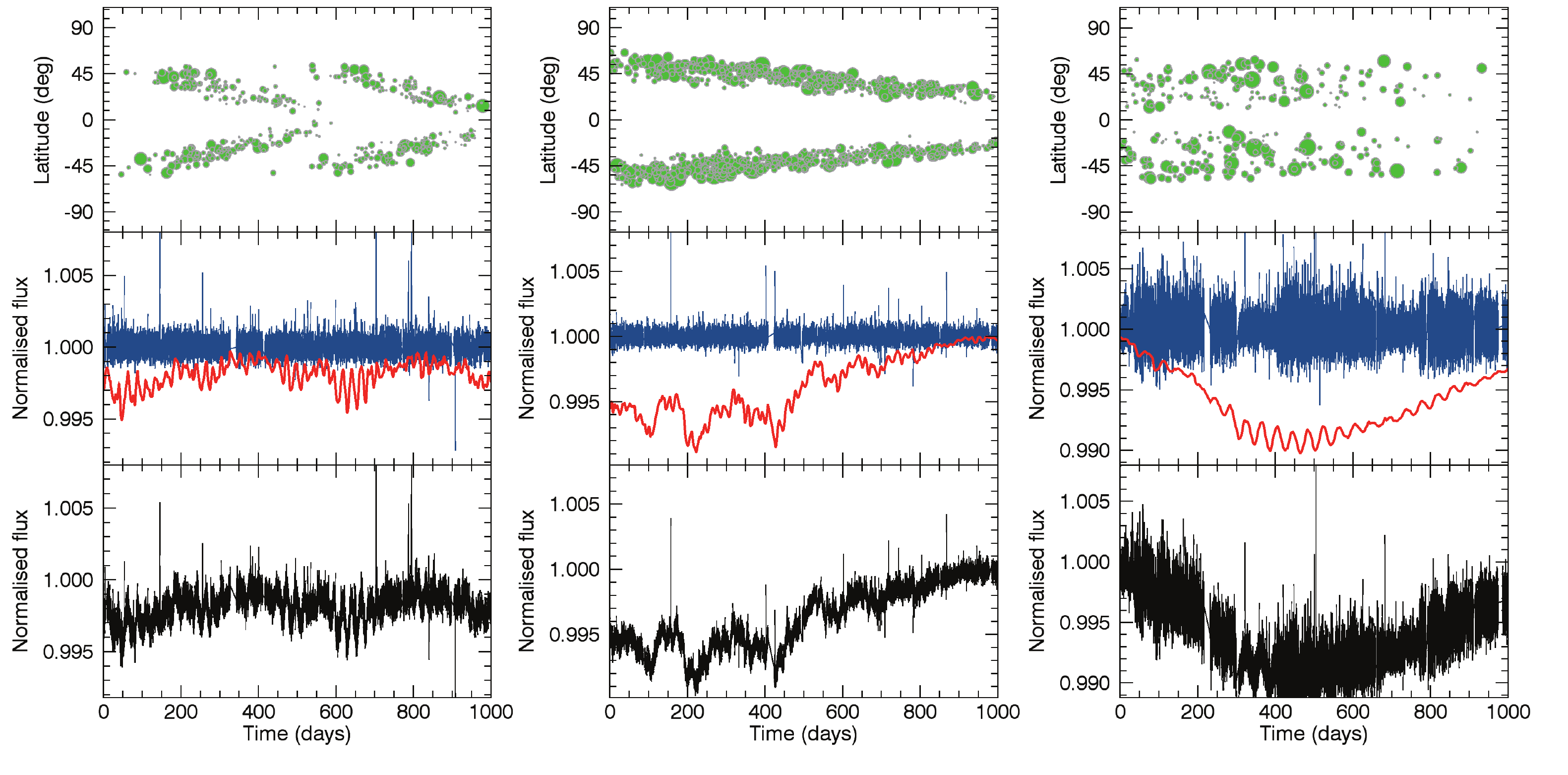}
  \caption{Example simulated light curves (left: no.\ 9, $P_{\rm eq} =
    23.5$\,days, $\tau=5.2\,P_{\rm eq}$, $\Delta\Omega/\Omega_{\rm
      eq}=0.33$; middle: no.\ 31, $P_{\rm eq} = 20.8$\,days,
    $\tau=1.1\,P_{\rm eq}$, $\Delta\Omega/\Omega_{\rm eq}=0.35$,
    right: no.\ 12, $P_{\rm eq} = 18.5$\,days, $\tau=1.6\,P_{\rm eq}$,
    $\Delta\Omega/\Omega_{\rm eq}=0.56$).  
    The top panel shows the emergence times and latitudes of the
    simulated active regions; the symbol size scales as the peak
    magnetic flux density. In the middle panel, the red dots show the
    simulated, noise-free photometric signal, the blue points the
    \emph{Kepler} PDC-MAP light curve the signal was injected
    into. The bottom panel shows the final, noisy light curve included
    in the blind exercise sample. The first two light curves have
    similar equatorial rotation periods and differential rotation
    shears, but the second has more rapid spot evolution. The
    third has both rapid spot evolution and very strong differential rotation, which appears as a random rather than
    butterfly-like spot distribution.}
  \label{fig:simlc}
\end{figure*}

\section{The hares: simulating the light curves}

This section describes the process by which the light curves used in the blind exercise were simulated. Note that we do not distinguish between `active regions' and `spots' in the simulation: each active region is effectively assumed to consist of a single dark spot whose area (or equivalently, contrast) evolves over time. 

\subsection{The spot emergence model}
\label{sec:spot}

The first task in simulating the light curves is to compute the time-dependent distribution of spots on the stellar surface. On the Sun, the rate and latitude of emergence of sunspots varies over an eleven year cycle, giving rise to what is known as the `butterfly pattern' first discovered by \citep{mau04,mau22}  (for a recent overview of the Solar cycle, see \citealt{hat10}). At the beginning of the cycle, spots emerge at $\theta\pm 35^\circ$. As the cycle progresses, the spots begin to emerge closer to the equator. During that time, the rate at which spots emerge first rises gradually, then decays again. After eleven years, the cycle repeats. The nature of this cycle is believed to be intrinsically linked to the solar dynamo \citep{ber05}.

Very little is known about the distribution and evolution of spots on stars other than the Sun. Doppler imaging, and Zeeman-Doppler imaging have revealed that, on rapidly rotating stars, star spots are not restricted to two distinct latitude bands. Instead, these stars appear to host spots at all latitudes on the stellar disc \citep{don+03}. Indeed, observations of the rapid-rotator AB\,Doradus over many years appear to show no pattern to the distribution of star spots. Rather, the observations show magnetic activity at all latitudes in all the observations
\citep{col+02,jef+07}.

The latitude distribution of star spots and its evolution over the star's activity cycle has a direct influence on the disk-integrated brightness fluctuations, and hence on the extent to which information about rotation and differential rotation can be recovered from the light curves. We therefore took care to  incorporate a wide variety of butterfly-like patterns in the simulated light curves, as well as some cases with no explicit relationship between spot latitude and the phase of the activity cycle. Our starting point was the code developed by \citet{lla+12}, itself derived from the work of \citet{mac+04}. This code, initially designed to simulate the emergence of magnetic flux on the surface of the Sun, allows for the creation of butterfly patterns of varying activity levels, cycle lengths, and distribution patterns. It was already used by \citet{lla+12} to simulate star spot distribution patterns in the context of \emph{Kepler} light curves, specifically light curves containing planetary transits. Their goal was to investigate whether stellar butterfly patterns could be revealed by looking for phase changes in the appearance of bumps in transit light curves of stars hosting a planet in a misaligned orbit. 

The names, ranges and distributions of the input parameters of the starspot generator are listed in the first five rows of Table~\ref{tab:pardist}. They are: the star's overall activity level $A$, relative to that of the Sun; the duration of the activity cycle, $C_{\rm len}$, in years; the amount of overlap between consecutive activity cycles $C_{\rm over}$, in years; and the minimum and maximum spot latitudes, $\theta_{\rm min}$ and $\theta_{\rm max}$.  For each simulated star, the model generates a list of spot emergence times, $t_k$, longitudes $\phi_k$ and latitudes $\theta_{k,0}$, as well as peak magnetic flux density $B^{({\rm max})}_k$ (in Gauss) at the centre of each spot. The activity rate, $A$, controls the overall rate of starspot and magnetic flux emergence relative to the Sun: a value of $A=1$ leads to a total number of spots and flux emergence, integrated over a 27 day period, similar to what is observed on the Sun. The individual parameter values used for each simulation are listed in Table~\ref{tab:parval}, and the top panel of Figure~\ref{fig:simlc} shows three examples of the resulting butterfly patterns.

In most cases, we simulate a butterfly pattern by allowing the latitude at which the spots emerge to decrease monotonically from $\theta_{\rm max}$ to $\theta_{\rm min}$ over the duration of the cycle in two latitude bands (one in the Northern and one in the Southern hemisphere), as illustrated in the top panel of Figure~\ref{fig:simlc}. At a given time $t$ in the simulation, the activity cycle phase is given by $\Phi=t \textrm{ mod } C_{\rm len}$, where $\Phi \in [0,1]$. {The spot emergence latitude as a function of phase is then given by $\theta(\Phi) = \theta_{\rm max} - (\theta_{\rm   max} -\theta_{\rm min})\Phi$. We also assume a latitude scatter in the spot emergence to match the observed Solar butterfly pattern \citep{hat10}. The spots are assumed to emerge at random longitudes, and the number of spots per unit time is defined to be inversely proportional to the square of the spot size \citep{sh94}. For 20\% of the cases, however, the spots were simulated to emerge randomly between $\theta_{\rm max}$ and $\theta_{\rm min}$. This is denoted by a $1$ (rather than a $0$) in the column $R$ in Table~\ref{tab:parval}. 

At the start of each simulation there are no spots on the stellar surface. To allow the simulation to reach a realistic `steady state', the butterfly patterns were simulated starting $N$ days before the start of the actual simulated light curves, where $N$ was a uniform random number between 200 and 400 days.

\subsection{Simulating the photometric signal}
\label{sec:simlc} 

\begin{table*}
  \centering
  \caption{Parameters of the simulated light curves (full version available online in machine-readable form). For a description of the last 3 columns, see Section~\ref{sec:pobs}}
  \label{tab:parval}
  \begin{tabular}{rrrrrrrrrrrrrrr}
\hline
$N$ & $A$ & $C_{\rm len}$ & $C_{\rm over}$ & $\theta_{\rm min}$ &
$\theta_{\rm max}$ & $R$ & $\sin i$ & $P_{\rm eq}$ & $\delta
\Omega/\Omega_{\rm eq}$ &
$\tau_{\rm dec}$ & KID & $P_{\rm obs}$ & $P_{\rm obs,min}$ & $P_{\rm obs,max}$ \\
& ($^a$) & (yr) & (yr) & ($^{\circ}$) & ($^{\circ}$) & ($^b$) & &
(day) & & ($P_{\rm eq}$) & ($^c$) & (day) & (day) & (day) \\
    \hline \hline
1 & 0.39 & 1.12 & 0.27 & 13 & 76 & 0 & 0.19 & 1.10 & 0.46 & 8.2 &
893468 & 1.51 & 1.26 & 1.77 \\
2 & 0.45 & 9.52 & 0.19 & 20 & 63 & 0 & 0.99 & 1.58 & 0.40 & 7.3 &
1163449 & 1.87 & 1.75 & 2.02 \\
3 & 1.34 & 1.54 & 1.53 & 33 & 74 & 0 & 0.34 & 17.04 & 0.00 & 1.02 &
1431091 & 17.04 & 17.04 & 17.04 \\ \hline
  \end{tabular} 
\medskip

Notes: ($^a$) $A$ is relative to solar; ($^b$) If $R$ is 0, spots follow a butterfly pattern, otherwise they are randomly distributed in latitude; ($^c$) A KID of 0 means that the noise-free light curve was used, while the entry `Sun' in this column means that the light curve is based on a segment of the Sun's observed total irradiance variations (see Section~\protect\ref{sec:sun} for details).
\end{table*}

After using the model described in Section~\ref{sec:spot} to generate star spot emergence and distribution patterns, the next step is to simulate the photometric signal of the spots on the rotating stellar surface. The names, ranges and distributions of the parameters controlling the light curve simulations are listed in the remaining rows of Table~\ref{tab:pardist}, and their individual values for each simulation in the remaining columns of Table~\ref{tab:parval}, while the remaining panels of Figure~\ref{fig:simlc} illustrate the process. The parameter ranges and distributions were chosen to be as realistic as possible while also covering (and hopefully slightly exceeding) the range of detectable signals. 

Each star was randomly assigned an inclination $i$ (defined as the angle between the stellar equator and the line of sight) and an equatorial rotation period $P_{\rm eq}$. The inclinations were drawn from a uniform distribution in $\sin^2 i$. This means that the simulated stars have a tendency to be seen closer to equator on, though all orientations are possible. For 90\% of the light curves, $P_{\rm eq}$ was drawn from a log uniform distribution between 10 and 50 days, while for the remainder it was drawn from a log uniform distribution between 1 and 10 days. This reproduces approximately the distribution of periods measured for \emph{Kepler} F, G and K main sequence stars by \citet{rei+13,mcq+14}.

We then implemented Sun-like differential rotation by setting the rotation rate of each spot to:
\begin{equation}
\Omega(\theta_k) = \Omega_{\rm eq} \left[ 1 -  \frac{\delta
    \Omega}{\Omega_{\rm eq}} \sin^2(\theta_k) \right], 
\end{equation}
where $\Omega_{\rm eq} = 2 \pi / P_{\rm eq}$. Two thirds of the light curves were given fairly strong differential rotation ($\delta \Omega/\Omega$ drawn from a log uniform distribution between 0.1 and 1) while the remaining third had no differential rotation.

We simulate the photometric signature of each spot using the very simple analytical model of \citet{aig+12}. The instantaneous relative flux drop caused by a single spot emerging at latitude $\theta_k$ and longitude $\phi^{(0)}_k$ is given by:
\begin{equation}
\delta F_k(t) = f_k(t) ~ {\rm MAX} \left\{\cos \beta_k(t), 0\right\},
\end{equation}
where $\beta_k(t)$ is the angle between the spot normal and the line-of-sight:
\begin{equation}
\cos \beta_k(t) = \cos \phi_k(t) \cos \theta_k \cos i + \sin \theta_k \sin i,
\end{equation}
$\phi_k(t) = 2 \pi (t-t_k) / P(\theta_k) + \phi^{(0)}_k$, and $f_k(t)$ is the flux drop that the spot would cause if located at the centre of the stellar disk. 

In physical terms, $f_k$ depends on the spot size, shape and contrast relative to the `clear' photosphere. However, in the present study, we set $f_k$ directly, using the following expression to allow for spot evolution: 
\begin{equation}
f_k(t) = \left\{ \begin{array}{ll}
f^{({\rm max})}_k \exp \left[ (t - t_k) \right] / (2 \tau_{\rm em}) & {\rm for}~t<t_k,~{\rm and} \\
f^{({\rm max})}_k \exp \left[ (t_k - t) \right] / (2 \tau) & {\rm for}~t \ge t_k. \end{array} \right.
\end{equation}
where 
\begin{equation}
f^{({\rm max})}_k \equiv 3 \cdot 10^{-4} A  B^{({\rm max})}_k /
\langle B^{({\rm max})}_k \rangle,
\end{equation} 
(the constant of proportionality was chosen to give rise to approximately Sun-like levels of variability for $A=1$), $\tau$ is the spot lifetime, which we assume to be the same for all spots on a given star, and $\tau_{\rm em}$ was set to either $\tau/5$ or 2 days, whichever was the longer. This two-sided behaviour was chosen to mimic the rapid growth and slower decay of spots observed on the Sun, while avoiding discontinuities in the light curves. A wide range of spot and/or active region lifetimes have been observed on the Sun and on other stars, ranging from a few days for the shortest-lived solar active regions to years for particularly active stars \citep{brad+14}. After some experimentation, we found that drawing $\tau$ from a log uniform distribution between 1 and 10 $P_{\rm eq}$ gave rise to light curve morphologies broadly similar to those observed by \emph{CoRoT} and \emph{Kepler}. 

Finally, the total observed flux for the star is simply
\begin{equation}
F = F_0 \left(1 - \sum_{k=1}^{K} F_k \right),
\end{equation}
where $F_0$ is the flux that would be observed in the absence of any spots. We simulated 1000 light curves in this manner, each lasting 1000 days with a cadence of 0.5\,hours (approximately equal to the \emph{Kepler} long-cadence interval). It is worthy of note that in our simulations the absolute continuum level is known and is used to normalise the entire light curve. In reality, the absolute continuum in \emph{Kepler} light curves is unknown. Since our simulated light curves have been consistently normalised, short stellar cycles could potentially be confused with rotation periods which may not be detected in real \emph{Kepler} photometry. Three examples of the resulting light curves are shown in the middle panel of Figure~\ref{fig:simlc}.

The simplified formalism we have adopted treats spots as point-like, ignoring the effects of spot shape, differential projection effects within the spot area, and overlapping spots. Limb-darkening is also ignored. These approximations could be problematic if one were trying to infer individual spot properties but not for the present study, where the main quantity of interest is the rotation period $P$. Another important advantage over a more physically realistic but more complex approach is computational speed: it takes only a few seconds to generate each simulated light curve. Finally, we note that our light curves are principally representative of stars exhibiting solar like activity. From the full \emph{Kepler} data set it is clear that a significant fraction of stars exhibit long-lived active regions that remain for many stellar rotations. Our simulated light curves do not reproduce such persistent spot patterns and so the findings reported in this exercise should not be extrapolated to such stars.

\subsection{Injection into PDC-MAP light curves}
\label{sec:inject}

In order to reproduce the noise properties of the \emph{Kepler} data, the signals as described in Section~\ref{sec:simlc} were injected into actual \emph{Kepler} light curves of quiet main-sequence stars. These were randomly selected from a sample of 7788 dwarf F, G, K, and M stars in which \citet{mcq+14} found no evidence of any periodic variability. Three such examples are shown in the middle panel of Figure~\ref{fig:simlc} (blue line), with the resulting, combined light curve shown in the bottom panel.  Of the 1000 simulated light curves, noise was added in this manner to the first 770. The remaining 230 were kept noise-free, in order to enable us to distinguish between the effects of noise and instrumental systematics, and more fundamental limitations of rotation period recovery in stellar light curves.

The approach adopted here does not reproduce the impact of the \emph{Kepler} systematics removal process on stellar signals. The PDC-MAP pipeline used to remove systematics in \emph{Kepler} data does affect astrophysical signals, albeit at a moderate level for timescales significantly shorter than a \emph{Kepler} quarter. The PDC-MAP retains signals with periods less than 3 days, whilst periods longer than 20 days are likely removed by the pipeline or distributed to other frequencies \citep{th13}. Ideally, we would inject the simulated signals into the \emph{Kepler} light curves before running them through the PDC-MAP pipeline, but this was not an option since the pipeline is not publicly available. 

\subsection{Solar test cases}
\label{sec:sun}
\begin{figure}
  \centering
  \includegraphics[width=\linewidth]{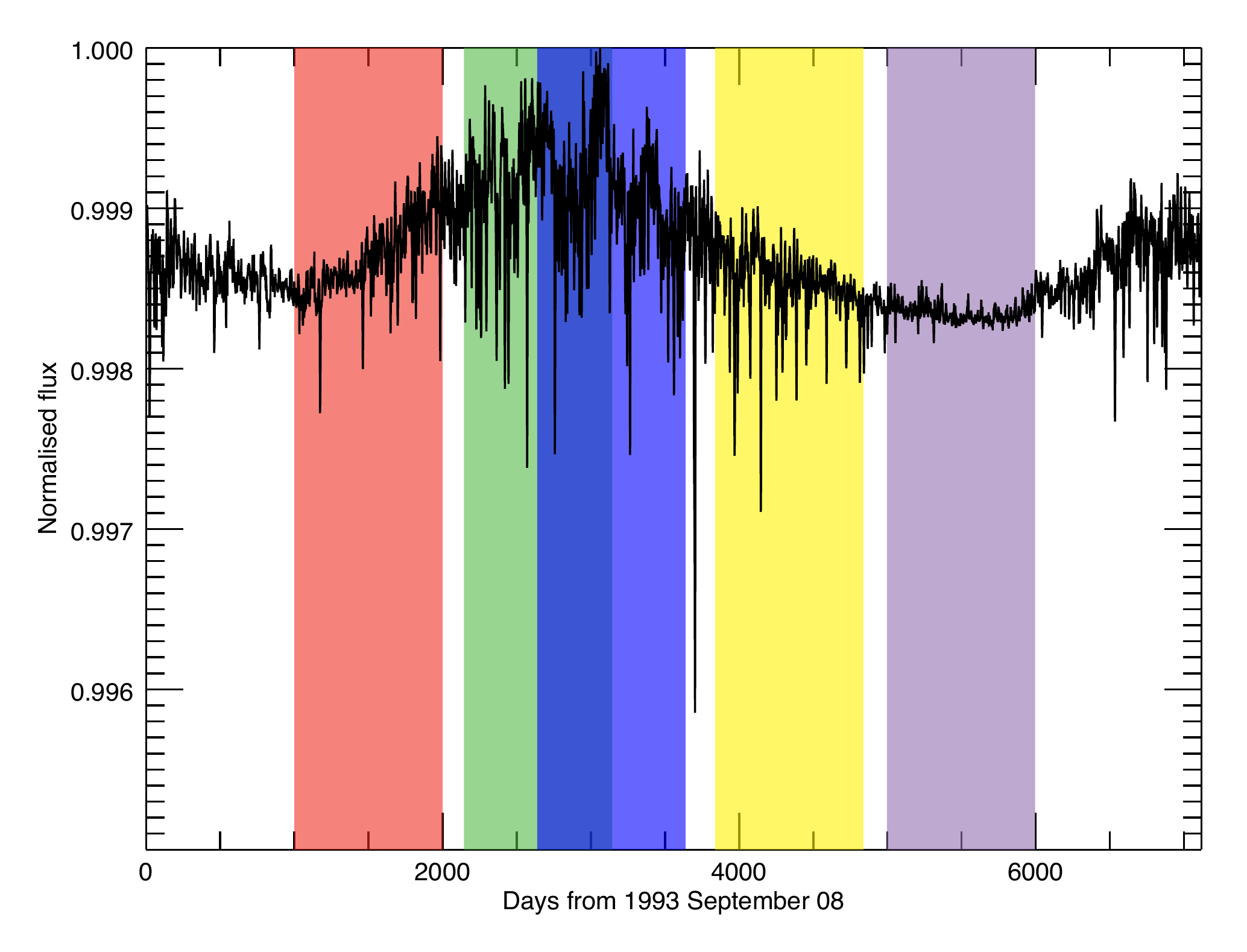} 
  \caption{Segments of total solar irradiance variations used to generate the last five light curves included in the set.}
  \label{fig:solar}
\end{figure}

To test the ability of the participating teams to recover the Sun's rotation period and differential rotation signal, we also included in the blind exercise set five light curves based on the composite total solar irradiance (TSI) time series  \citep{fl98,fro00,fro06}, taken during the last solar cycle, and thus based on data from the VIRGO instrument on the SOHO spacecraft. The data were downloaded from the PMOD/WRC website ({\tt http://www.pmodwrc.ch/}), and rebinned to the same cadence as the simulated light curves described in Section~\ref{sec:simlc}. We then selected five 1000-day long segments spanning the rising phase, maximum, decaying phase and subsequent minimum of the last complete Solar activity cycle, as illustrated in Figure~\ref{fig:solar}. 

\section{The hounds: Measuring rotation and differential rotation}

\begin{figure}
  \centering
  \includegraphics[width=\linewidth]{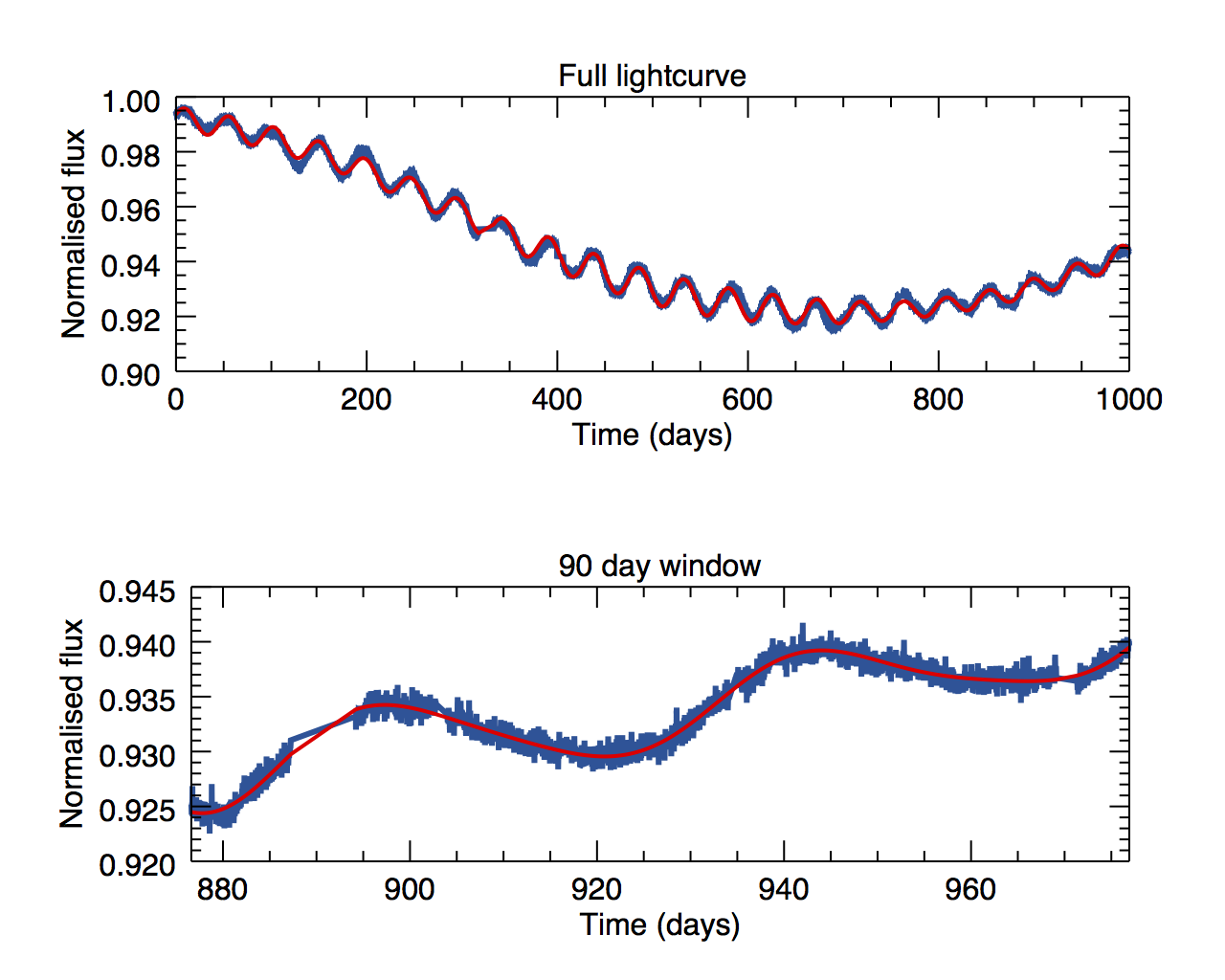}
  \caption{Example multi-component sinusoidal fits obtained by the G{\"o}ttingen/Reinhold team for light curve 9 (see text for details). The top panel shows the results for the full light curve, and the bottom panel for a randomly selected 90-day segment. The data are shown in blue and the fits in red.}
  \label{fig:reinhold_ex}
\end{figure}

\subsection{The G{\"o}ttingen / Reinhold team}
This team analysed the light curves using the Lomb-Scargle (LS) periodogram of the full (1000-day) light curves, in a standard pre-whitening approach as described in \cite{rei+13}. This approach consists of locating the most significant peak in the LS periodogram, subtracting the best-fit sinusoid with that period from the data, computing the LS periodogram of the residuals, and repeating the process 5 times. Once the five periods have been identified, a global sine fit is performed according to
\begin{equation}
  y_{\rm fit} = \sum_{k=1}^5 a_k \sin\left(\frac{2\pi}{P_k}\,t-\phi_k\right)+c,
\end{equation}
where $c$ is a global offset, and $P_k$, $a_k$ and $\phi_k$ are the period, the amplitude and phases of the $k^{\rm th}$ component. The best parameters were found using $\chi^2$-minimisation, resulting in five refined period values for each light curve. The procedure is illustrated for an example light curve in Figure~\ref{fig:reinhold_ex}.

The LS periodograms were computed over the period range $0.5$ to $100$ days, and the first period, $P_1$, was selected within this range. To minimise the number of cases where only the first harmonic of the true rotation period was detected, $P_1$ was compared to the remaining periods $P_{k}$ by computing $|2 P_1 - P_{k}|/2 P_1$. If this quantity was $<0.05$, the highest-peak period $P_k$ satisfying this relation, which is most likely to be the true rotation period, was chosen as the main reported period.

The presence of a second period $P_2$ adjacent to $P_1$ is indicative of either differential rotation, spot evolution, or a combination of both effects. If there was a period $P_2$ satisfying the relation 

\begin{equation}  
 0.01 \leq |P_1-P_2|/P_1 \leq 0.30
\end{equation}
this period was also reported, in order to test whether the interval between $P_1$ and $P_2$ could be used as a measure of differential rotation. If any $P_3$, $P_4$ and $P_5$ also satisfied this criterion, the largest and smallest of these were reported as $P_{\rm min}$ and $P_{\rm max}$.

The analysis described above was also applied to a randomly selected 90-day window from each light curve, this time searching for periods up to 45 days to account for the much shorter time span. After visual inspection of the light curves, periodograms and periods returned by the two approaches (full light curve versus 90-day segment), the periods extracted from the full light curve analysis were deemed to be more reliable in most cases ($>90$\,\%). The only exceptions were cases where the main period derived from the analysis of the 90-day window was less than 10 days: the analysis of the full light curve appeared less sensitive to short periods.  Thus, the final reported periods $P_1$ (and $P_2$ if measurable) were chosen from the analysis of the full light curve, except for these short-period cases, where the periods derived from the 90-day segments were reported.

The method was applied automatically to all light curves, and a value was reported for all cases where the main period was $<100$\,days, leading to a total of 840 detections, with at least two periods reported in 545 cases. This detection rate is larger than in \citet{rei+13}, where it was $\sim 60\%$. The difference is probably due to the longer period cutoff used here (100 instead of 45-days), which enables the detection of longer periods but also, potentially, spurious signals caused by instrumental effects or long-term spot evolution. To quantify this, 100 light curves were inspected visually, in order to evaluate the expected false-positive rate. About 20 of the visually inspected light curves had detected periods without clear counterpart in the light curve, leading to a predicted false-positive rate of approximately 20\,\%.

\subsection{The G{\"o}ttingen / Nielsen team}

This team also used the LS periodogram, but worked with shorter segments of data, equivalent to individual \emph{Kepler} quarters (90 days). The analysis method was based on \citet{nie+13} and was identical to that study in all respects, apart from the number of quarters analysed. Each 1000-day time series was divided into $10$ artificial quarters of roughly $90$ days. The LS periodogram was computed for each quarter using the method described in \citet{fra+95}. For each quarter, the highest peak was recorded, provided that it was at least higher than $4$ times the white noise level. This level was computed based on the root mean square (RMS) of the time series, as described in \citet{kb95}. The LS periodogram was computed over the range $0.5$--$90$ days, but cases where the period of the highest peak was $>45$ days were discarded, in an effort to minimize spurious detections from uncorrected instrumental variability or other sources of red noise.

For each star, the median absolute deviation of the periods measured from each quarter was then computed. Any star exhibiting a median absolute deviation above 1 day was discarded. This ensures that only stars with long lived variability are accepted, thus increasing the likelihood that the reported periods are real. This procedure is illustrated in Figure~\ref{fig:pvq}. The remaining stars were required to show at least 7 periods within 3 median absolute deviations of the median. We note that, if differential rotation is present, this would select only light curves where the signal was dominated by spots located at a specific latitude (or rotating at a specific rate) over most of the time span of the data. For stars that satisfy the above criteria, the median of the detected periods was reported as the final period, and the median absolute deviation as an estimate of the uncertainty on that period. This resulted in 158 detections, a somewhat larger detection rate than in \citet{nie+13}, but the latter searched the entire \emph{Kepler} sample, which included early-type, giant, and old stars, all of which are not expected to lead to period detections.

\begin{figure}
\centering
\includegraphics[width = \linewidth]{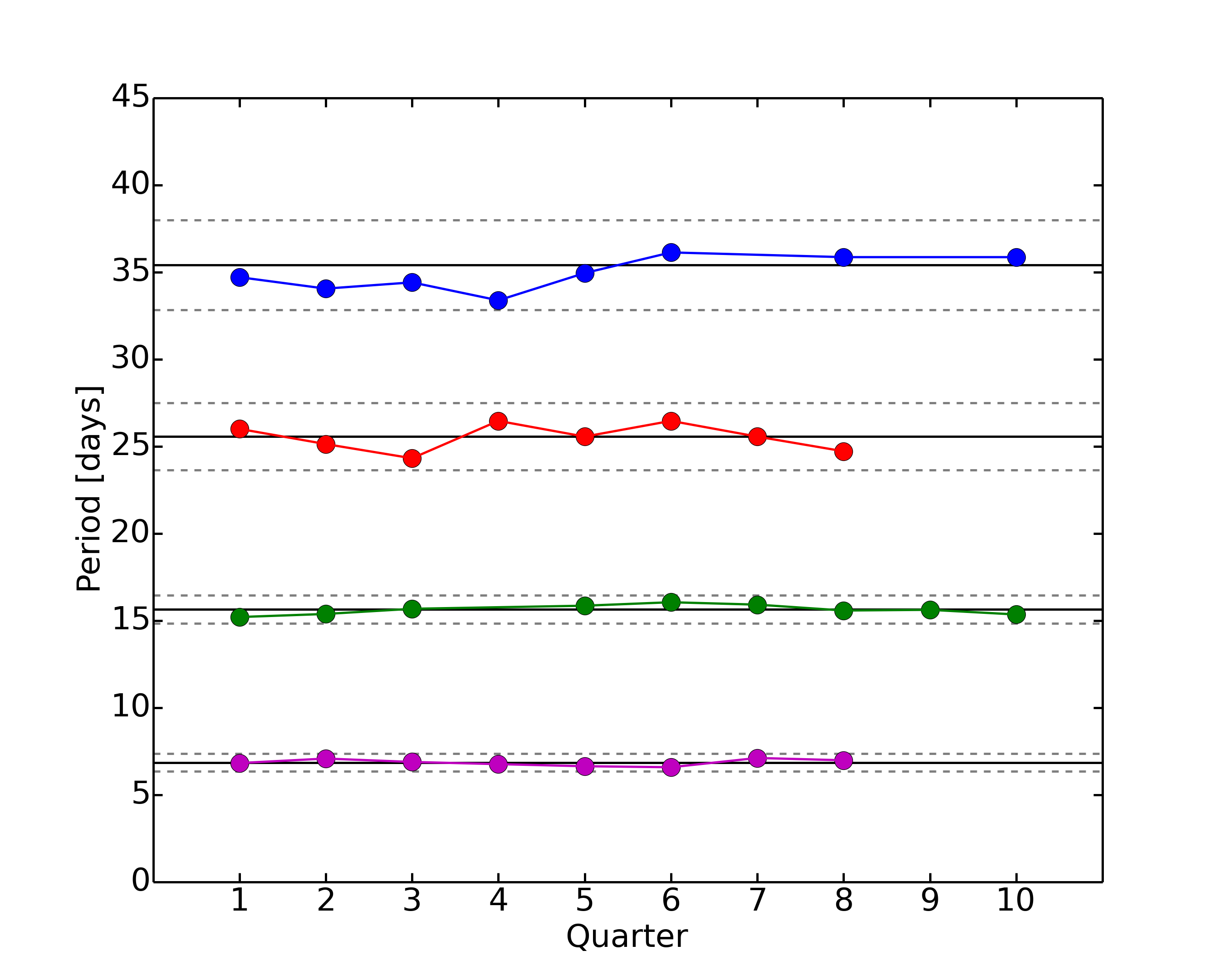}
\caption{Illustration of the selection criteria used by the  G{\"o}ttingen / Nielsen team. The periods measured in each quarter  for four stars are shown in different colours, while black  horizontal lines denote the corresponding median values (solid) and  $\pm3$ times the median absolute deviation (dashed). Only periods  lying between the dashed lines were retained.}
\label{fig:pvq}
\end{figure}

An attempt was also made to use the measured periods to estimate the differential rotation, if present. (This part of the analysis method was not included in previously published work). For the stars satisfying the selection criteria described above, a linear trend was fit to the measured periods versus quarter number. However, in almost all cases, no significant trend was found: the linear fit did not significantly reduce the scatter in the measured periods. Therefore, only the median period and uncertainty were reported.

\subsection{The Tel Aviv team}

\begin{figure}
  \centering
  \includegraphics[width=0.8\linewidth]{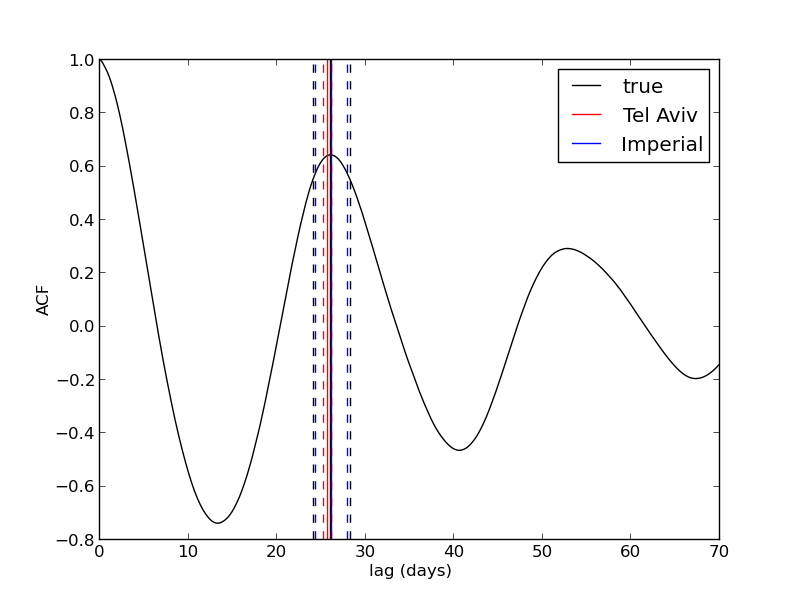}
  \caption{Autocorrelation function of light curve 9, as computed by the Imperial team, with the period ranges reported by the Tel Aviv and Imperial teams shown by the vertical red and blue lines, respectively. The vertical black lines show the `true' period range, computed as described in Section~\ref{sec:pobs}.}
  \label{fig:acf_ex}
\end{figure}

This team analysed the light curves using the autocorrelation function (ACF) method first introduced by \citet{mcq+13a}. The ACF is essentially a measure of the degree of self-similarity of the light curve at different time intervals, or lags. Periodic or quasi-periodic signals with period $P$ in the light curve give rise to an oscillating ACF with regularly spaced peaks at lags $P$, $2P$, \ldots, modulated by a decaying envelope resulting from correlated noise, spot evolution and/or differential rotation. 

The light curve pre-processing and implementation of the ACF method used in this exercise were identical to that applied to the full \emph{Kepler} sample, as described in \cite{mcq+14}, except that a) a pre-processing step was implemented to remove very long-term variations, and b) no attempt was made to select reliable periods, manually or automatically. 

The simulated light curves contain very long-term variations caused by changes in overall spot coverage. These were not present in the \emph{Kepler} light curves analysed in \citet{mcq+14}, because the PDC-MAP processing removes them. It was necessary to remove them here, as they would otherwise disrupt the identification of the rotation period from the ACF.  The light curve was smoothed using a non-linear filter \citep{ai04}, consisting of a median filter followed by a boxcar filter, with 5000 and 2500-point windows respectively. The resulting smoothed light curve was subtracted from the original, before computing the ACF. The reported period and associated uncertainty were then measured by identifying up to 4 regularly spaced peaks in the ACF, and fitting a straight line to peak lag versus peak number. Figure~\ref{fig:acf_ex} shows the ACF and period range obtained for light curve 9.

In previous studies using the ACF, reliable periods were selected either visually \citep{mcq+13a,mcq+13b} or automatically using a criterion based on the relative height of the first ACF peak, stellar effective temperature and detected period \citep{mcq+14}. The latter could not be applied here, since effective temperatures were not available. As selection based on visual inspection would have run the risk of introducing human bias into the results of the exercise, no selection was done and a period was reported for every light curve.

\subsection{The Imperial team}

This team also used the ACF, but with a slightly different implementation. In particular the pre-processing of the light curves was as follows. First, a Savitzky-Golay filter \citep{sg64} with third order polynomials over three month sections of the light curve was applied to smooth long term variability whilst preserving the underlying signal. The light curves were then normalised by dividing them by their median and subtracting unity. Any gaps were linearly interpolated across and Gaussian noise added. Finally, a median filter of width 11 time steps (equivalent to 5.5 hours) was applied, followed by a boxcar filter, also of width 11 time steps, to smooth out the high-frequency variability. 

The ACF of the resulting light curves was then computed. Peaks in the ACF were deemed significant if higher than 0.15. If no peaks above this value were present, no period was reported for that light curve. In most cases, the first peak was taken to correspond to the rotation period. However, when the second peak was higher than both the first and the third peaks, the second peak was selected as the primary peak. Any primary peaks corresponding to a rotation period of less than two days were excluded, as these are particularly strongly affected by correlated noise in the light curve, and their locations are thus uncertain. Primary peaks beyond 50 days were also excluded, as signals longer than this would be removed by the PDC-MAP pipeline in \emph{Kepler} data, and any remaining signal beyond 50 days would typically be due to instrumental effects. This resulted in detections for 970 of the 1005 simulated light curves.

Once the primary peak was identified, up to four subsequent peaks within 10\% of integer multiples of the primary one were also identified. A rough estimate of the uncertainty on the lag of each peak was estimated by measuring the full width at half-maximum (FWHM) of each of the selected peaks, the uncertainty on the lag of each peak was then taken as ${\rm   FWHM} / (2 \sqrt{2 \ln 2} )$. A straight line was then fit to peak lag versus peak number, and the slope of the fit and the resulting uncertainty were adopted as the period and period error, respectively.  Figure~\ref{fig:acf_ex} shows the ACF and period range obtained for light curve 9.

\subsection{The Natal team}

\begin{figure}
  \centering
  \includegraphics[width=0.9\linewidth]{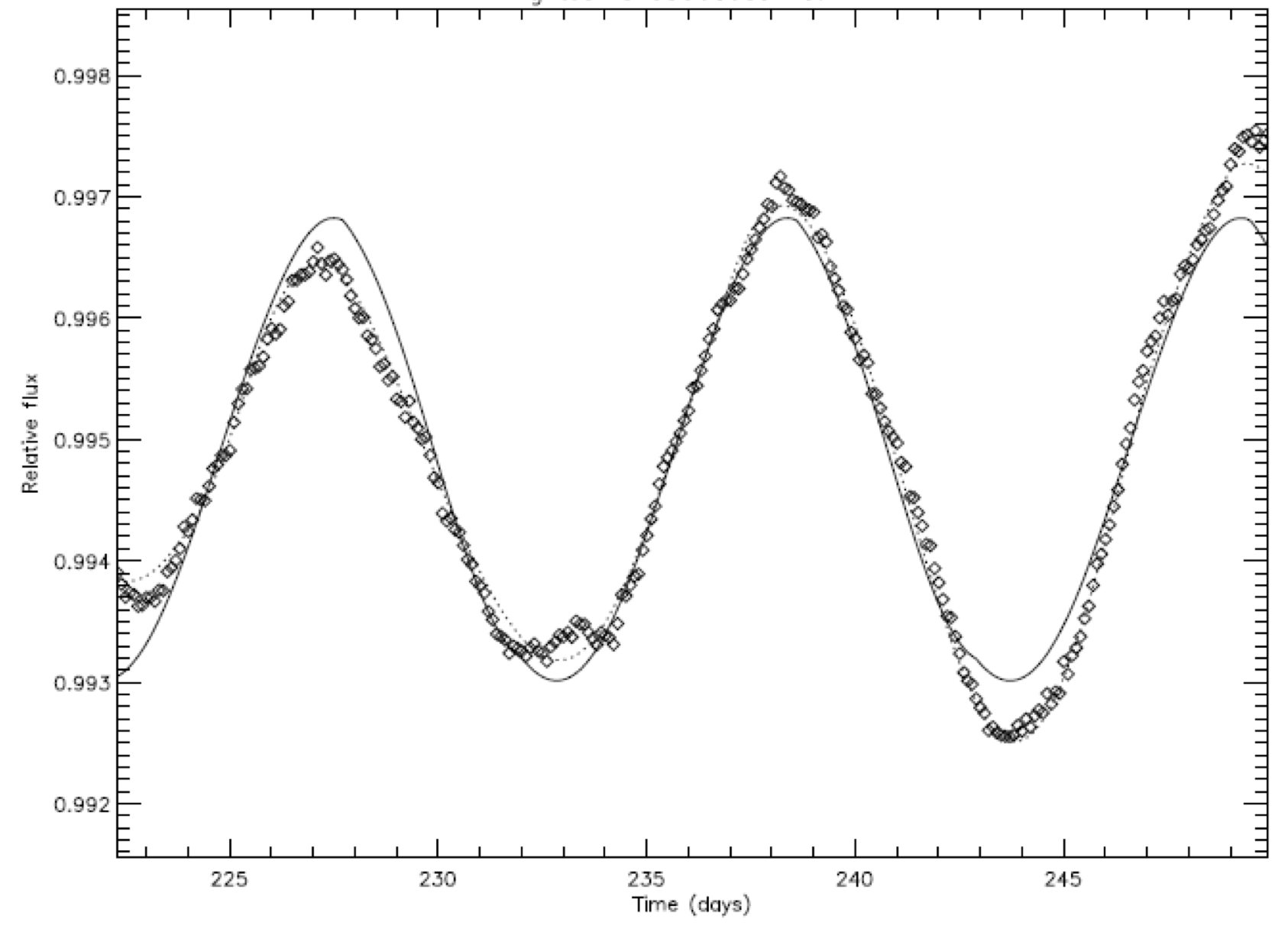}
  \caption{Illustration of the two-spot modelling carried out by the Natal team. This shows, for light curve 506, the segment showing the strongest evidence for differential rotation. The data are shown by the open lozenges, the best-fit obtained with differential rotation by the dotted line, and the best fit obtained with rigid rotation by the solid line. The minimum and maximum periods reported by the Natal team for this light curve were 10.77 and 11.62 days, respectively, while the `true observable' minimum and maximum periods, as defined in Section~\protect\ref{sec:pobs}, were 10.62 and 11. 39 days.}
  \label{fig:natal_dr}
\end{figure}

This team used a combination of periodogram, ACF and spot modelling to analyse the light curves, following the procedure described in \citet{lan+14a}. As this process involves a number of manual steps, and the spot modelling is computationally expensive, only the last 500 light curves (from no.\ 506 onwards) were analysed by this team. 

First, the light curves were detrended following the method adopted by \citet{dem+13}. This involves identifying discontinuities using a simple algorithm, and removing long-term trends by fitting a third-order polynomial to each quarter, similar to the recipe of \citet{bas+11}. When applied to raw \emph{Kepler} data, this procedure gives results that are fairly similar to those obtained by subtracting a linear combination of the PDC-MAP co-trending basis vectors from the raw data.

Next, a generalised Lomb-Scargle periodogram \citep{zk09} was computed for each light curve. The highest peak of the periodogram was used as a preliminary estimate of the star's rotation period (with no additional selection criteria), as an input to the spot modelling.

The next step used the ACF to identify the light curves with the most stable rotational signal, which are the best candidates to estimate the mean rotation period as well as differential rotation. Working with a small sample of active stars observed by \emph{CoRoT}, \emph{MOST} and \emph{Kepler}, \citet{lan+14a} were able to measure what they interpreted as a differential rotation signal when the relative height of the first peak in the ACF was around 0.6--0.7.  The present hare-and-hounds exercise represents a unique opportunity to test this conclusion. Thus, the spot modelling was also applied to cases where the first ACF peak was lower than this value (down to 0.5). However, owing to time limitations, it was only applied to 303 of the possible 500 light curves.

The remaining 303 light curves were fit using a two-spot model, as described in detail in \citet{lan+14a}.  To allow for spot evolution, the time series were cut into segments of equal duration $T$, during which spots can be considered to remain stable. Each segment is then fit with a two-spot model. The angular rotation rate of the two spots were initially set to the value given by the periodogram analysis mentioned above, but were allowed to vary independently within $\pm15\%$ of this value. The inclination of the star was initially fixed to $60^{\circ}$, as it is very degenerate.  The spot model was fit using {\sc mpfit}, an {\sc idl} implementation of the Levenberg-Marquart algorithm for non-linear least squares, which allows for some or all of the parameters to be constrained within certain bounds. As this is a local optimiser, the choice of initial parameters can strongly affect the results, and not all segments result in acceptable fits, even for short $T$. Starting with $T$ equal to the light curve duration, shorter and shorter intervals were considered (by gradually increasing the total integer number of intervals) until the fit became acceptable (see below).

The segments were then fit again with a 2-spot model without differential rotation, i.e.\ forcing the two spots to share the same rotation period. The initial period of both spots was set to that of the first spot as derived from the differential rotation fit (and all the other parameters' initial values were also taken from the differential rotation fit). For each segment, the relative improvement in the fit between the differential and solid rotation models was computed using the Bayesian Information Criterion (BIC). Any segment where $\Delta~{\rm BIC}>2$ in favour of the differential rotation model were visually examined, and the team manually selected the segment that showed the clearest evidence for differential rotation, using it to report the final minimum and maximum periods. For each light curve, this visual examination was done for a range of segment durations $T$, and the acceptability of the fit was evaluated at the same time. Typical values for the final segment duration are of the order of 1.5 to 2 rotation periods. 

As there is a strong degeneracy between stellar inclination and several of the spot parameters (mainly latitude and size), the inclination was not explicitly varied in the fits. Instead, it was initially fixed at $60^{\circ}$. If no acceptable fit was found with this inclination, progressively lower values were tried (in steps of $5^{\circ}$), until an acceptable fit was obtained.

Finally, for the two cases deemed to show the strongest evidence for differential rotation (light curves 506 and 513), a Markov Chain Monte Carlo (MCMC) analysis was used to obtain posterior distribution over the spot parameters.  

\begin{figure*}
  \centering
  \includegraphics[width=0.8\linewidth]{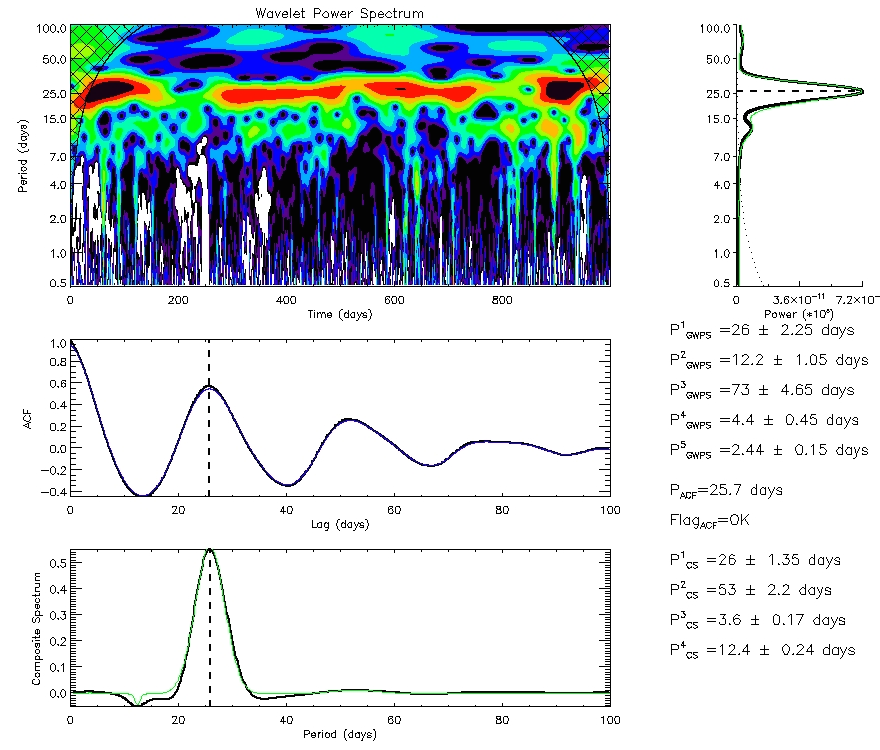}
  \caption{Example graphical output produced by the CEA team for light curve 9. The top left panel shows wavelet transform of the detrended light curve. The top right panel shows this transform collapsed onto the period axis, while the middle and bottom panel show the light curve ACF and the composite spectrum obtained by multiplying the wavelet spectrum with the ACF. In the last 3 panels, the thick black line shows the data, thin green lines show a fit to this composed of multiple Gaussians, and the thin purple line shows the smoothed version of the ACF used to identify the ACF peaks. The various periods identified from the different spectra are reported in the bottom right corner of the figure}
  \label{fig:cea_ex}
\end{figure*}

\subsection{Non-blind teams}

Two more teams took part in the exercise after the content of the simulations had been made public. These teams did analyse the light curves blindly, in the sense that they did not check the content of individual light curves. Nonetheless, they had access to considerably more information about the range and distribution of the  simulated parameters, and also had benefited from the prior experience of the other teams, whose results were discussed before these two additional teams took part.

\subsubsection{The CEA team}

This team used the ACF, an analysis based on a wavelet decomposition of the light curves, and the product of the two. The three results thus obtained are then compared to ascertain that the detected period is not method-dependent. This analysis is very similar to the one used by \citet{gar+14}, and identical to Cellier et al.\ (in prep.), and is illustrated in Figure~\ref{fig:cea_ex} for one example light curve.

The pre-processing of the light curves used in this exercise is based on a simplified version of the tools developed to treat the \emph{Kepler} simple aperture photometry light curves \citep{gar+11}. In particular, the light curves were smoothed by convolving them with a 5000-point ($\sim 100$ days) triangular kernel. As the signals were injected into the PDC-MAP rather than raw light curves, no jump correction or quarter-by-quarter detrending was applied. The light curves were then rebinned by a factor of 4 to speed up the analysis.

Their analysis then consisted of three parts. Firstly, the Morlet wavelet decomposition of the light curve is computed, yielding a time-frequency spectrum. This decomposition is then projected onto the period axis to obtain the Global Wavelet Power Spectrum (GWPS), which is similar to a Fourier power spectrum but with degraded resolution. As shown by \citet{mat+10}, the GPWS is less susceptible to detecting harmonics of the true rotation period compared to a standard periodogram. Moreover, the time-frequency spectrum can be used to check if the periodic signal is present throughout the light curve, or is caused by a localised data artefact. The main peaks of the GWPS are then fitted with Gaussian functions and the period corresponding to the highest peak is stored as $P_{\rm rot,GPWS}$. The half-width at half-maximum of the fitted Gaussian is taken as the uncertainty on this period, this also takes into account any possible contribution from differential rotation.

Secondly, the ACF of the light curve is computed and smoothed. The smoothing length is determined by the strongest peak in the periodogram of the ACF. Then the period corresponding to the first ACF peak is stored as $P{\rm rot,ACF}$. Following \citet{mcq+13a}, the team checked for repeated peaks in the ACF and regular patterns in the light curves to ensure that the detection is genuine. As it is difficult to calculate uncertainties from the ACF, they did not give an error on this period. The primary goal of computing the ACF period was to validate the GPWS period detection.

Finally, a composite periodogram was obtained by multiplying the ACF with the GWPS. This is done to boost the peaks that are present in both the ACF and the GWPS and to reduce the amplitude of the peaks that only appear in only one of the two methodologies. They call the result the Composite Spectrum (CS). As for the GWPS, the main peaks of this CS are fitted with Gaussian functions and the period corresponding to the highest peak is returned as $P_{\rm rot,CS}$, while the uncertainty is given by the half-width at half-maximum.

For each light curve, the three periods, $P_{\rm rot,GWPS}$, $P_{\rm rot,ACF}$, and $P_{\rm rot,CS}$ were then compared. If the periods were consistent to within 10\% of each other, the detection was considered confirmed. The final rotation period was then taken to be $P_{\rm rot,GWPS}$ with the associated uncertainty. This last, automatic validation step is the main difference between the present analysis and the approach used in \citet{gar+14}, which used a much more time-consuming, systematic visual check.


\begin{figure*}
  \centering
  \includegraphics[width=\linewidth]{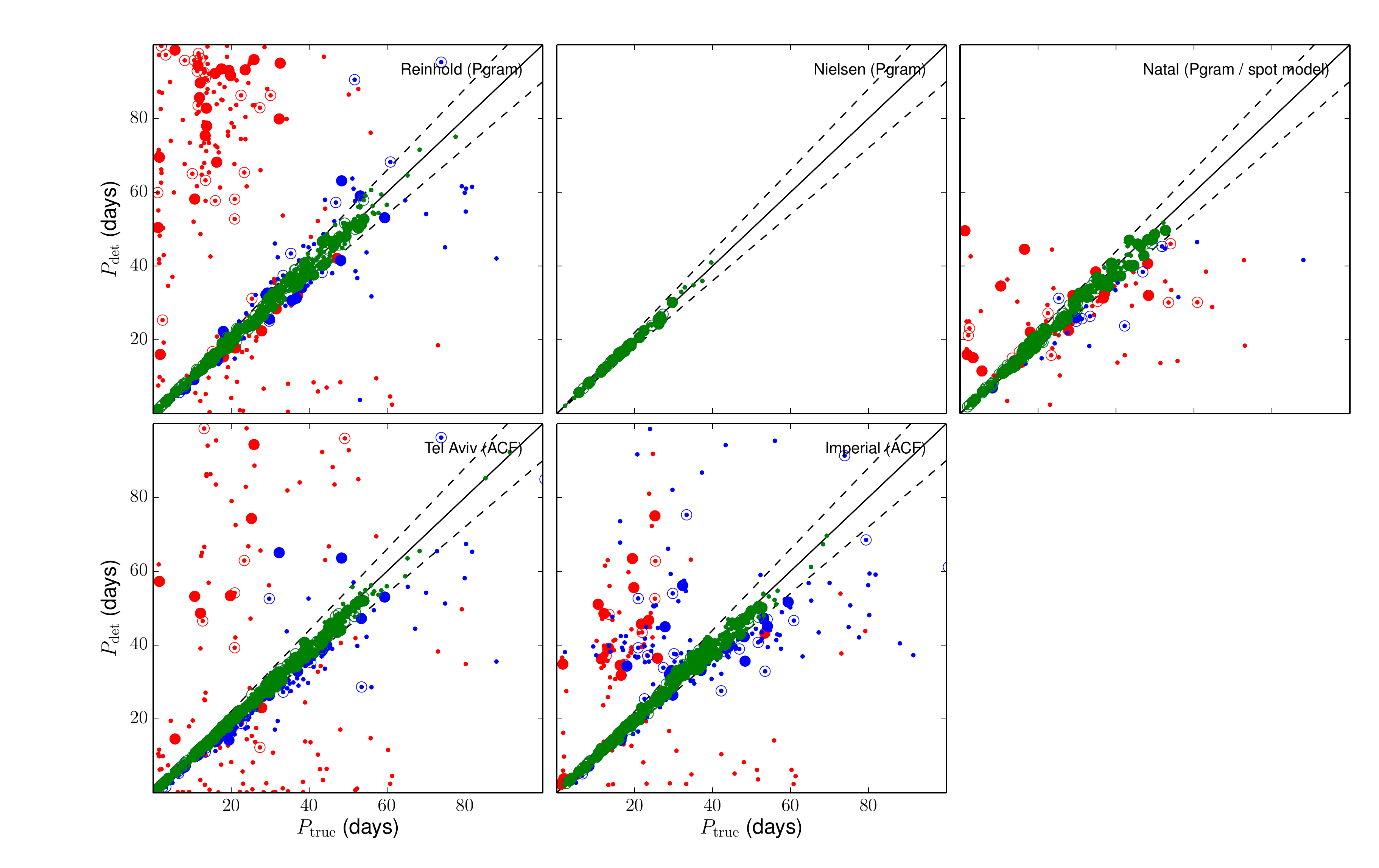} 
  \caption{Detected versus simulated periods for each team. Green, blue and red symbols correspond to `good', `ok', and `bad' detections, respectively (see text for details). The larger symbols correspond to the noise-free light curves. The solid and dashed lines mark the 1-to-1 and $\pm 10$\% correspondence between simulated and detected periods. The unfilled symbols correspond to stars with a cycle length less than three years, which may be confused with a long rotation period in the analysis performed by the teams}
  \label{fig:per_res}
\end{figure*}

\subsubsection{The Seattle team}

The Seattle team used a similar approach to the period finding as the G{\"o}ttingen/Reinhold team, based on the normalised Lomb-Scargle Periodogram. Each light curve was first detrended to remove the longer time-scale variations and residual \emph{Kepler} systematics. A boxcar smoothing algorithm, with a width of 3000 data points ($\sim 82$ days) was applied to the light curve and the result subtracted from the original, before computing the LS periodogram. The LS periodogram was not oversampled in this analysis.

Since a smoothing kernel of 82 days was applied, only periods shorter than this were considered significant. Manually examining the light curves led to a conservative upper limit of 75 days and a lower limit of 0.25 days for the range of allowed periods. Rather than selecting the highest amplitude peak in the LS power ($P_x$) as the most likely period, the Seattle team used a modified criterion. The highest amplitude peak of the quantity $R_x=P_x \sqrt{\nu}$ was selected, where $P_x$ was the normalised LS power as a function of frequency, and $\nu$ the corresponding frequency. This had the effect of requiring peaks at longer periods (shorter frequencies) to have larger amplitudes. The standard deviation over all frequencies of $R_x$ was also computed, and peaks were considered ``significant'' and thus reported for this exercise if they were above this standard deviation.

Only a single peak was measured in this modified LS analysis, and no checks for alias periods were performed. After discussions with the participating (blind and non-blind) teams, and inspection of the simulation input parameters, the Seattle team did not directly measure a second period from the 1000-day light curves. Instead, the spread of the LS power about the peak was estimated, which was observable as multiple peaks near the primary period, or power distributed over a range of periods with no distinct sub-peaks. The LS spread was estimated by first boxcar smoothing the $R_x$ curve with a kernel of 5 frequency bins, and then least-squares fitting the smoothed $R_x$ with a Gaussian function. The standard deviation of the Gaussian, $\sigma_{Rx}$, was reported, as it does correlate with the range of periods simulated via differential rotation for some light curves. However, non-zero LS spread was also observed for objects having no differential rotation but with several starspots emerging at different longitudes (or equivalently rotation phase) throughout the 1000-day light curve. 

\begin{figure*}
  \centering
  \includegraphics[width=0.8\linewidth]{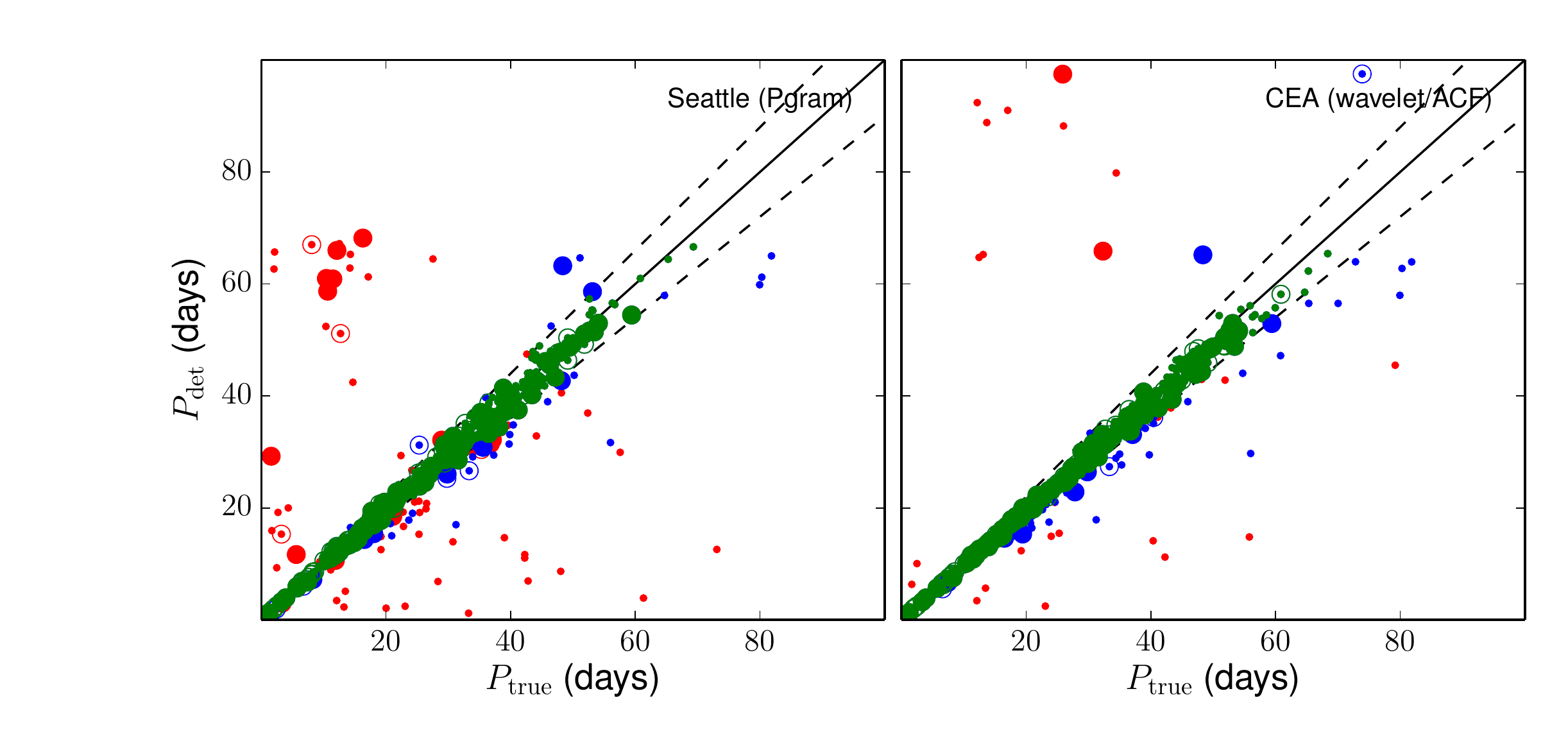} 
  \caption{Detected versus simulated periods for the two teams who analysed the light curves in non-blind mode. Lines and symbols are the same as in Figure~\ref{fig:per_res}.}
  \label{fig:per_res_new}
\end{figure*}

\section{Results}

In this Section, we evaluate how well the different teams were able to measure the `overall' rotation period, as well as the amount of differential rotation, in each light curve. First, we must define the `ground truth' to compare the different team's results to.

\subsection{Observable periods}
\label{sec:pobs}

For each star, the simulations define an equatorial and a polar rotation period, plus a minimum and maximum spot latitude. Coupled with the differential rotation prescription, this corresponds to a minimum and maximum rotation period that could be present in the data. However, these values cannot necessarily be compared directly to the periods measured from the light curves. If the activity cycle duration is significantly longer than the light curve duration, the simulated spots may cover only part of the allowed latitude range. Furthermore, the signal will tend to be dominated by the largest active region, and may not be representative of the periods of spot population as a whole.

To enable a more adequate comparison between the simulated and measured periods, we proceed as follows. For each light curve, each spot was assigned a weight proportional to its peak effective area (see Section \ref{sec:simlc}). The `overall', minimum and maximum periods were then defined as the median, 10$^{\rm th}$ and $90^{\rm th}$ percentiles of the weighted periods. These values are reported in the last 3 columns of Table~\ref{tab:parval}.

\subsection{Recovery of overall rotation period}

\begin{table*}
  \centering
  \caption{Overall rotation period detection results. For the noisy and noise-free simulated light curves, the 3 columns shows the percentage of cases for which each team reported a value, and the percentage of reported values that were `good' and `ok', respectively (see text for details). For the solar cases, we report numbers rather than percentages.}
  \label{tab:per_res}
  \begin{tabular}{lrrrrrrrr}
\hline
   Method & \multicolumn{3}{c}{Noisy} & \multicolumn{3}{c}{Noise-free} & \multicolumn{2}{c}{solar} \\
   & \% det & \% good & \% ok & \% det & \% good & \% ok & No.\ det & No.\ ok \\
\hline \hline
   \multicolumn{9}{c}{Blind teams} \\ \hline
   Reinhold (Periodogram) & 82 & 67 & 76 & 87 & 65 & 77 & 4 & 2 \\
   Nielsen (Periodogram) & 16 & 100 & 100 & 15 & 100 & 100 & 0 & 0 \\
   Natal (Periodogram  / spot model) & 27 & 69 & 76 & 72 & 80 & 85 & 5
   & 2 \\
   Tel Aviv (ACF) & 100 & 68 & 80 & 100 & 75 & 90 & 5 & 4 \\
   Imperial (ACF) & 95 & 71 & 87 & 98 & 73 & 88 & 5 & 4 \\ \hline
   \multicolumn{9}{c}{Non-blind teams} \\ \hline
   Seattle (Periodogram) & 68 & 81 & 88 & 75 & 84 & 90 & 0 & 0 \\
   CEA (Wavelets / ACF) & 78 & 88 & 95 & 82 & 92 & 99 & 2 & 2 \\ \hline
  \end{tabular}
\end{table*}

Depending on the team and on the individual light curves, in some cases a single period estimate was reported along with an uncertainty, in others two or more period estimates, and in others still a period range. Where a team reported more than one value without explicitly identifying which was the dominant one, we used the mid-point of the range of reported periods as the `detected' period. Where a single uncertainty value was reported rather than a range, we defined the detected range as the detected period $\pm 1 \, \sigma$. 

A period detection was considered `good' if the true and detected period were within 10\% of each other, `ok' if there was any overlap between the simulated and detected period ranges, and `bad' otherwise. Figures~\ref{fig:per_res} and \ref{fig:per_res_new} show, for the blind and non-blind teams respectively, for each team, the simulated versus reported periods, for the noisy and noise-free light curves. Table~\ref{tab:per_res} reports, for all the teams, the fraction of light curves for which a period was reported, and the fraction of light curves for which the reported period was `good' or `ok' according to the definition above.

Of the five teams who participated in a genuinely blind fashion, three (G{\"o}ttingen/Reinhold, G{\"o}ttingen/Nielsen, and Natal) identified the `main' period as the highest peak in the Lomb-Scargle periodogram, or variants thereof, but with different inputs (length of light curve segment considered, range of periods searched) and selection criteria (coherence across multiple quarters, visual examination). As a result, the fraction of the light curves for which these teams report a detection varies from $\sim 15$\% (G{\"o}ttingen/Nielsen) to $\sim 80$\% (G{\"o}ttingen/Reinhold). As one might expect, increasing the number of detections comes at the expense of reliability: between the most and the least conservative of these (G{\"o}ttingen/Nielsen and G{\"o}ttingen/Reinhold), the percentage of `good' detection drops from 100 to $\sim 75$. The remaining two blind teams (Tel Aviv and Imperial) used the auto-correlation function (ACF) to measure the `overall' period. Both report even higher detection fractions ($>95$\%), while retaining fairly good reliability (just over 70\% of the detections were `good' and close to 90\% were `ok'). 

Overall, these results are very encouraging: for the range of periods and amplitudes simulated, overall rotation periods can be recovered with high completeness, and high reliability, using either the periodogram or the ACF. If one is interested in a fully automated method that will give results for almost all light curves while retaining high reliability, an ACF-based method appears the most suitable. A periodogram-based approach can give even better results when the rotation signal is strong, at the expense of some completeness.

We now turn to the results of the non-blind teams.  The Seattle team falls somewhere in between the two G{\"o}ttingen teams in terms of completeness and reliability,  reporting results for $\sim 70\%$ of the light curves, of which $\sim90\%$ were `good'. This is essentially a consequence of slightly different selection criteria. The CEA team reported periods for almost 80\% of all light curves, of which close to 90\% were `good'. These represent the best combination of completeness and reliability among all the teams. Bearing in mind that this was not a completely blind analysis, it nonetheless suggests that a combination of wavelet and ACF analysis, together with dedicated pre-processing, can lead to even better results. 

\subsubsection{Nature of the `bad' detections}

For all teams except the CEA one, `bad' detections, particularly at the longer end of the period range searched, occur at approximately the same rate among both noisy and noise-free light curves. This implies that these bad detections are not predominantly caused by residual systematics in the \emph{Kepler} PDC-MAP light curves.  

Almost all of the cases where one or more team reported a `bad' period that was significantly longer than the `true' value have short spot lifetimes. These are cases where the detected signal is not due to rotational modulation, but to the evolution of the star's overall spot-coverage. Although this evolution is not in fact periodic, it can give rise to some apparently quasi-periodic signal in the light curves. 

The lower rate of this kind of bad detection among the CEA results may be due to the light curve pre-processing approach used by that team. On the issue of pre-processing, it is important to note that the present exercise did not test the full impact of the PDC-MAP processing on stellar rotation signals, since the signals were injected into the processed light curves. We note again, that in our simulated light curves the entire light curve is normalised to the absolute continuum which is not known for real \emph{Kepler} light curves. This introduces an additional issue for the teams, who may have confused the stellar cycle with stellar rotation. In reality this would not be an issue when dealing with real \emph{Kepler} data because the normalisation between quarters is unknown. The teams that either restricted their analysis to shorter periods, or those teams that fit out the longer variations (in effect treating our light curves more like real \emph{Kepler} data) performed better than those teams who did not. The unfilled circles in Figures \ref{fig:per_res} and \ref{fig:per_res_new} show those stars with a stellar cycle length less than three years. A considerable number of the detections for these stars overestimated the rotation period of the star, suggesting that the short cycle length can indeed be confused with the rotation period of the star.

All teams also reported a smaller number of spurious detections that were significantly shorter than the true values. Interestingly, these occur only in the noisy light curve. A more detailed inspection of these cases revealed that the detected signal was present, albeit at a low level, in the `quiet' light curves into which the simulated signals were injected. Thus, these detections may not be entirely spurious, though it is not possible to determine whether the detected signal in those cases was genuinely due to rotational modulation of star spots. 

\begin{figure*}
  \centering
  \includegraphics[width=0.9\linewidth]{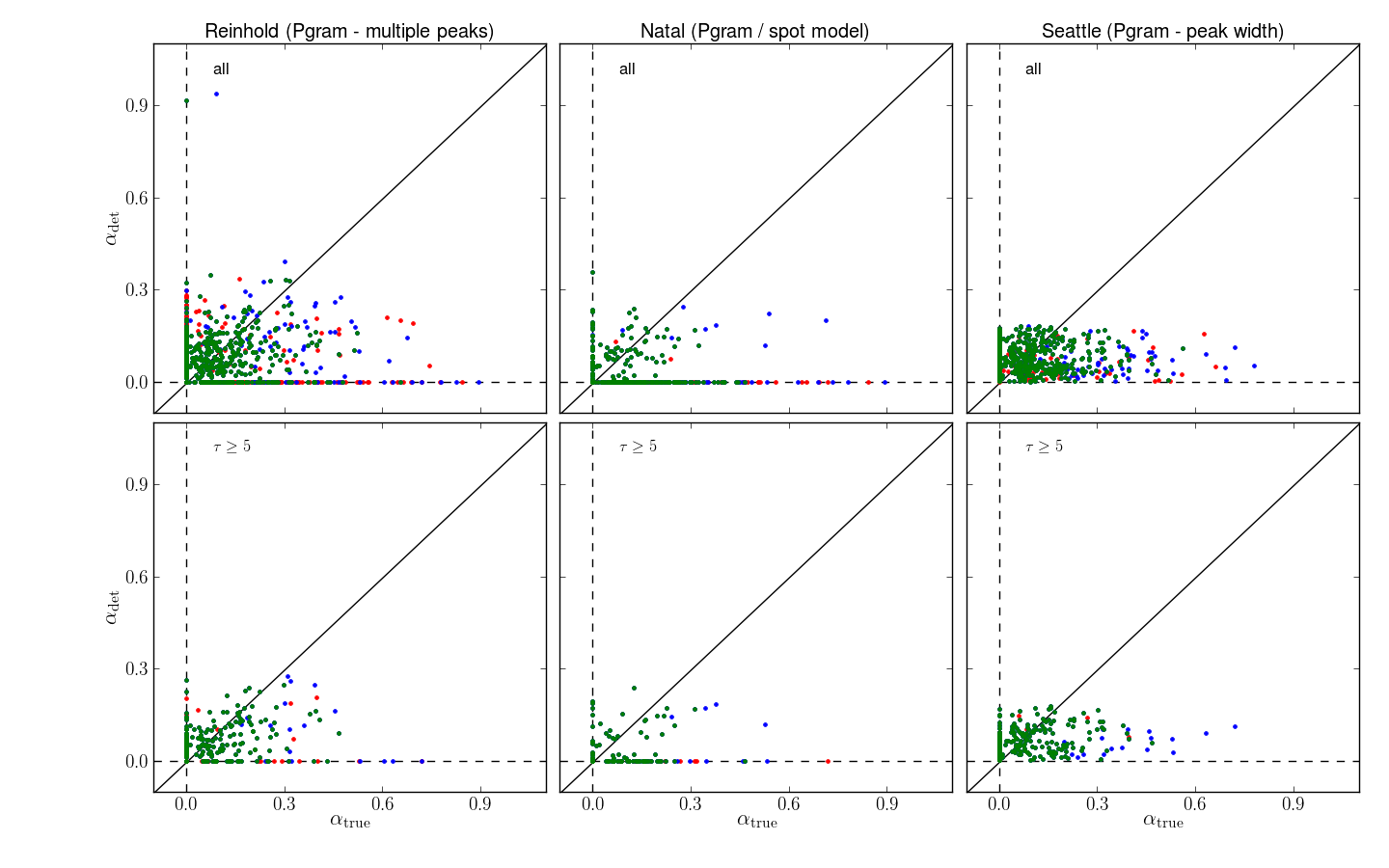}
  \caption{Measured versus simulated differential rotation shear ($\alpha \equiv (P_{\rm min} - P_{\rm max}) / P$) for the three teams which specifically searched for differential rotation signals. The colour-coding of the points is the same as in Figure~\ref{fig:per_res}. The top row shows all the light curves each team reported results for, while the bottom row shows only cases with spot evolution timescales $\tau > 5 P$.}
  \label{fig:diffrot_res}
\end{figure*}

\subsubsection{Solar cases} 

For the solar cases, a detection was considered `ok' if the reported period range overlapped with the interval 25--30 days. This interval covers the expected solar synodic rotation periods for latitudes 0--$45^{\circ}$), the approximate latitude range within which sunspots tend to be located.

Two of the teams using the periodogram recovered the solar period in 2 of the 5 solar cases, both the teams using the ACF did so in 4 of the 5 solar cases, and the CEA team recovered it in 2 of the 5 cases. This suggests, once more, that the ACF method may perform slightly better than the periodogram for Sun-like stars which are neither fast rotators nor very active. Additionally, we note that, on average, the measured period decreases as one progresses in the solar activity cycle, which is an encouraging sign for the detectability of differential rotation for \emph{Kepler} stars having shorter activity cycles. Previous studies of the TSI have shown that the true rotation period of the Sun was detectable only during phases of minimum activity during the 11 year solar cycle when the modulation was dominated by faculae \citet{lan+03,lan+04}. A re-analysis of the TSI using an autocorrelation approach was carried out in \citet{lan+14b} where they reached similar conclusions. 

\subsection{Recovery of differential rotation}

Not all teams explicitly searched for differential rotation signals. For those that did, Figure~\ref{fig:diffrot_res} shows a comparison of the simulated versus measured differential rotation signal, quantified by the rotational shear $\alpha$, which we define as $\alpha \equiv (P_{\rm max}-P_{\rm min}) / P$, following \citet{rei+13}, where $P$ is the `overall' or mean rotation period. To compute the simulated value, $\alpha_{\rm true}$, shown on the figure, we use the `observable' period range defined in Section~\ref{sec:pobs}, but the results are similar when using the absolute maximum and minimum periods defined by the spot latitude range and the value of $\Delta \Omega$ used in the simulations.

The three teams whose results are shown in Figure~\ref{fig:diffrot_res} used very different methods: identification of multiple individual peaks in the periodogram, spot modelling, and measurement of the spread of the power in the LS periodogram about the strongest peak. Strikingly, none of these methods enables an unambiguous detection of the differential rotation signal. All teams reported non-zero shear in numerous cases where the simulated shear was zero, and vice-versa. There are, of course, cases where each team reported a non-zero shear and the simulations did include differential rotation. Although there is a lot of scatter, there is some tendency for those to cluster around the one-to-one line, at least for the G{\"o}ttingen/Reinhold and Natal teams, particularly when considering only `good' period detections (green points), but the correlation is weak.  

Rapid spot evolution can lead to light curve morphologies that closely resemble those caused by differential rotation. In an attempt to disentangle between the effects of spot evolution and differential rotation, we plotted in the bottom panel of Figure~\ref{fig:diffrot_res} only the cases with relatively long spot lifetimes $\tau \ge 5 P$. There are fewer `bad' or `ok' period detections (red and blue points) in that subsample, but the picture in terms of recovery of differential rotation is not substantially different. It may be that differential rotation signals can only be identified reliably in cases where $\tau$ is even larger, but the present blind exercise did not include enough such cases to test that hypothesis.

We also checked the results of the other teams, who reported a period range or error bar, even when they did not do so with the specific intention of making a differential rotation measurement, but the results were similar: there was little or no correlation between the simulated shear and the period range or period uncertainty. We also checked for any correlation between simulated shear and a number of ACF properties measured by the Tel Aviv team, which could conceivably be related to differential rotation, such as the width of the ACF peaks, or the gradient in this width for consecutive peaks, but to no avail. 

\begin{figure*}
  \centering
\includegraphics[width=\linewidth]{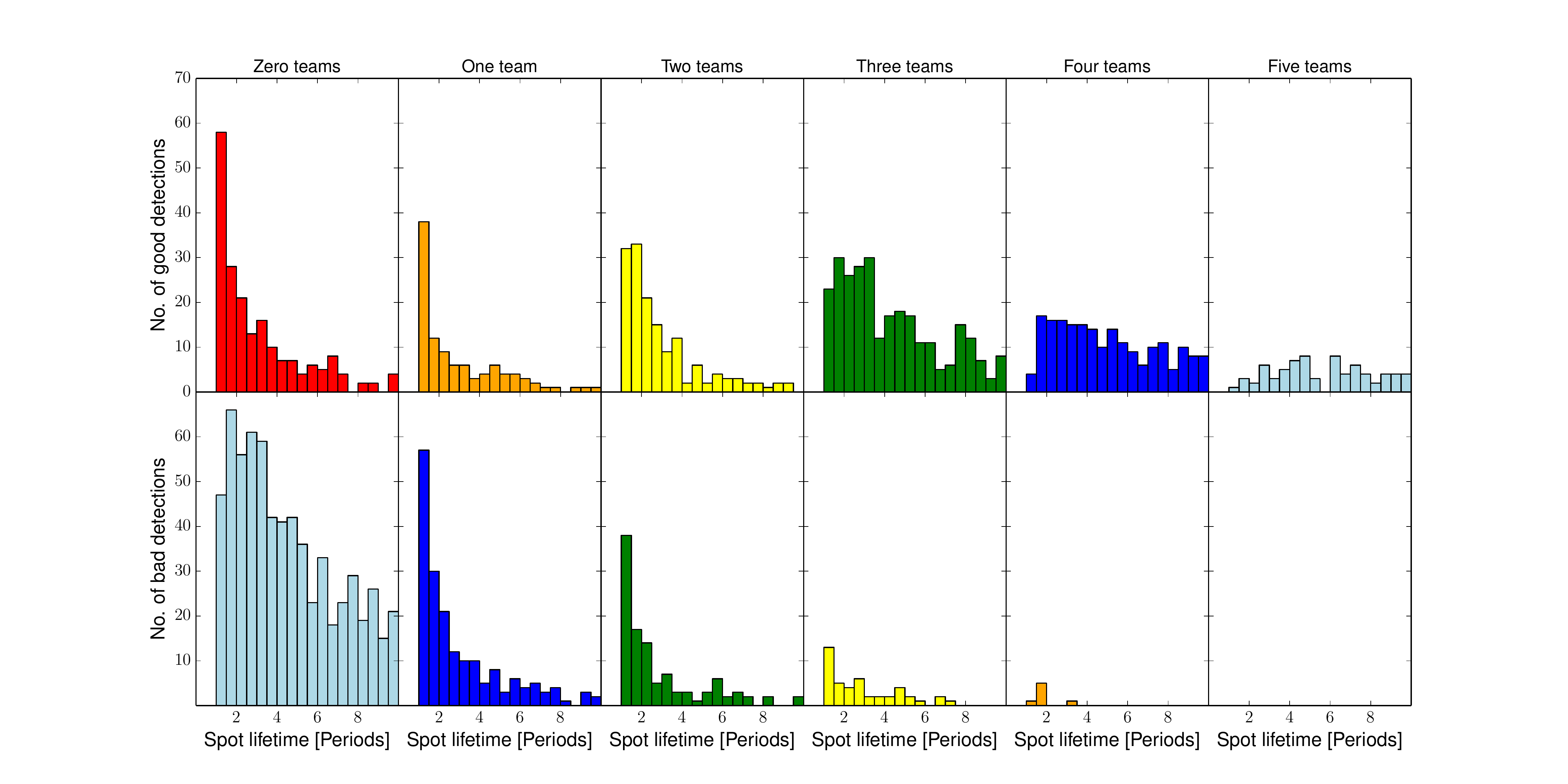}
\includegraphics[width=\linewidth]{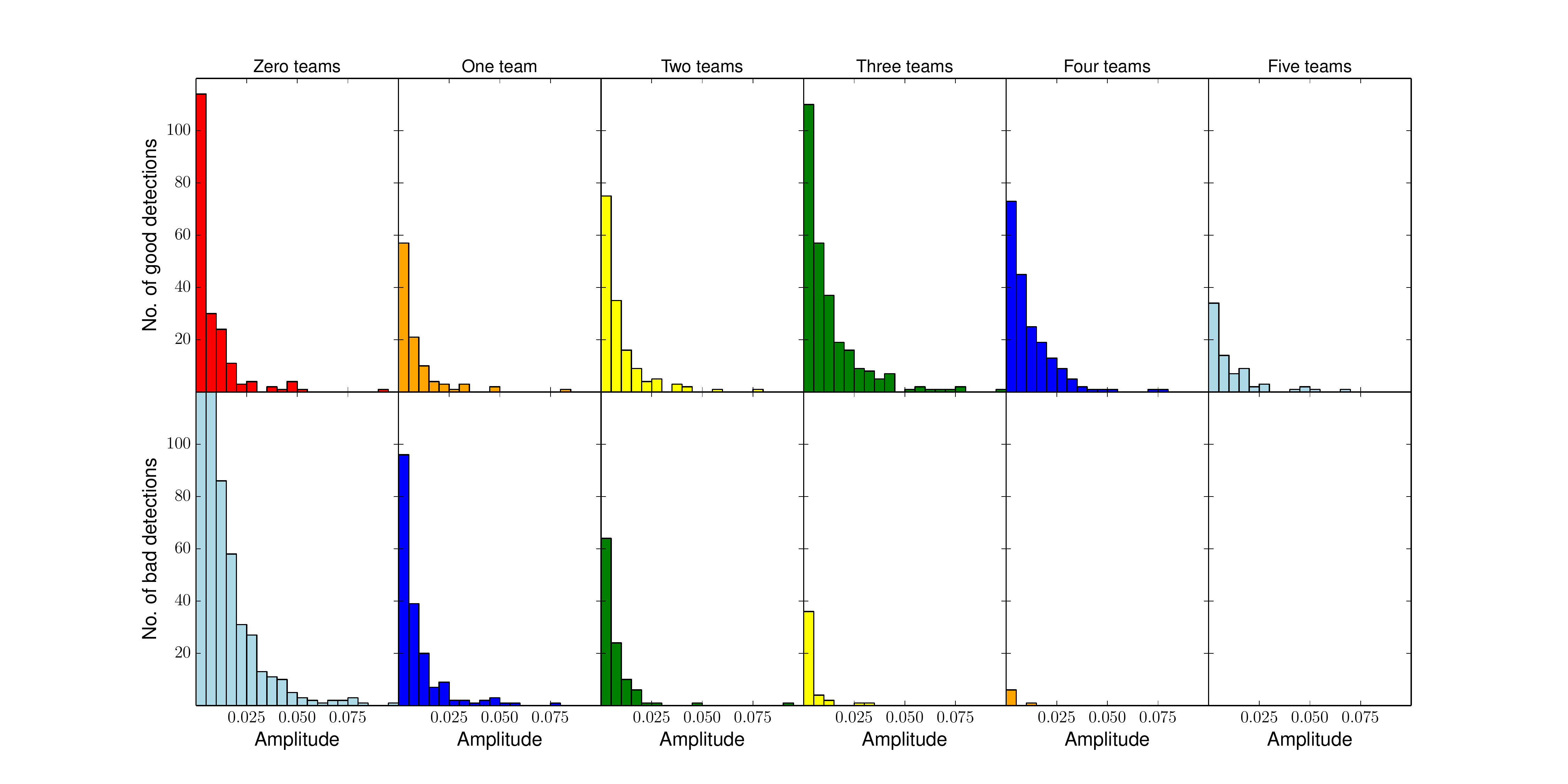}
 \caption{Histograms showing how the value for spot lifetime (top) and amplitude of variability (bottom) gave rise to consistently good or bad detections. For both parameters, the top row of histograms shows the number of teams with good detections and the bottom row the number of bad detections. } 
  \label{fig:no_teams}
\end{figure*}

\section{Summary and discussion}

We have carried out a blind exercise to test the recoverability of `typical' rotation and differential rotation signals from \emph{Kepler} light curves. Seven teams analysed up to 1004 light curves containing simulated starspot signals including activity cycles, differential rotation and spot evolution, using a variety of methods including variants on the Lomb-Scargle (LS) periodogram, autocorrelation function (ACF), wavelet power spectra and combinations thereof. Five of the teams took part in genuinely blind fashion, while the last two teams analysed the light curves after their content had been made public, though they did not explicitly use this information in their analysis. 

The results of the exercise for the recovery of the `overall' rotation period are positive: the periods reported by the different teams were within 10\% of the `true' period between 70 and 100\% of the time. There was a clear trade-off between completeness and reliability: one team -- G{\"o}ttingen/Nielsen, using the periodogram with a stringent stability criterion -- obtained 100\% `good' detections, but reported results for only 27\% of the cases, while another -- Tel Aviv, using the ACF method with no selection criterion whatsoever -- reported results for 100\% of the case, but only $\sim 70\%$ of these were `good'. Arguably, the best combination of completeness and reliability was obtained by the CEA team, using a combination of multiple detrending as well as period-search methods (ACF and wavelets). However, as this team took part after the end of the genuinely blind phase of  the exercise, their results cannot be compared directly to those of the other teams. 

The results are much more worrying for differential rotation. There seems to be very little correlation between the injected and recovered differential rotation shear. In numerous cases, a non-zero shear was reported by several teams where the simulated shear was zero, and vice-versa. The origin of this problem, which persists whatever the method used, is not entirely clear. Spot evolution is likely to be playing a role, as it can mimic differential rotation signals in the light curves and periodograms, but the problem persists even in cases where the simulated spots had relatively long lifetimes ($\ge 5$ rotation periods). Further tests with fewer, larger, longer-lived active regions are needed to see if differential rotations signals can be recovered more reliably in more active stars -- which have typically been the focus of differential rotation studies. In the mean time, the results of the present exercise suggest that differential rotation measurements based on periodogram analysis or spot modelling of the full-disk light curve alone should be treated with considerable caution.

It is worth noting that all of the teams who searched for differential rotation signals did so by looking for multiple periodicities present in the light curves at the same time. An alternative, which may yield more robust results, would be to look for a drift in the mean period over an activity cycle \citep{sca+13,mat+14}. However, this requires light curves lasting at least a significant fraction of the activity cycle. Furthermore, it implicitly assumes the existence of a Sun-like butterfly pattern, which may or may not be common in other stars. 

In the remainder of this section we comment on a few points which were brought up by the exercise and seem worthy of further discussion.

\subsection{What do we actually mean by rotation period?}

One question which had to be addressed in the course of the exercise, and which is less trivial than it may seem, is: what do we mean by a `correct' rotation period measurement for a star with differential rotation? We chose to address this in a statistical, rather than a strictly physical sense, by considering the distribution of spot rotation periods present in the simulations, weighted by the spot size, and comparing that to the reported period(s). This enabled us to evaluate the effects of noise, spot evolution, inclination angle of the star and the period measurement method used on the results. However, it is important to bear in mind that both the `true' and the reported period ranges are affected by other factors such as the distribution of latitudes at which spots emerge and the specific differential rotation prescription, which would be unknown \emph{a priori} in a realistic scenario.

\subsection{Main factors affecting overall period sensitivity and reliability}
 
Figure~\ref{fig:no_teams} shows the number of teams who reported `good' and `bad' detections, as a function of the (observable) spot lifetime and the peak-to-peak amplitude of the noise-free light curve. For the sake of consistency, only the results of the blind teams are shown, although the picture does not change appreciably when the results of the non-blind teams are included. As one might expect, both the sensitivity and the reliability of the period search, when evaluated across all the teams, decreases for short-lived spots, smaller amplitudes, and longer periods.  

\subsection{When are unambiguous detections of differential rotation feasible?}
While the differential rotation results of the exercise were on the whole disappointing, there were a few cases where individual teams did report a range of periods that closely matched the simulated range -- such a case is illustrated for the Natal team in Figure~\ref{fig:natal_dr}, for example. This naturally begs the question: if differential rotation can be reliably measured under some circumstances, can we identify what those circumstances are? We searched for this by comparing the spot lifetimes, median number of spots present at any one time, overall periods and activity levels for the stars where the reported shear was close to the simulated one, to those where it was not. Unfortunately, we were unable to identify, within the parameter space covered by this exercise, a `safe' region where differential rotation can be reliably measured. Even if future studies were to reveal that differential rotation can be detected reliably for -- say -- spot lifetimes longer than 10 rotation periods and stars dominated by only 2--3 active regions at any one time, this information would be useful only if coupled with a reliably means of establishing, from the light curve itself, that these conditions apply.

To our knowledge, the only differential rotation studies to explicitly account for spot evolution so far are the work of \citet{fro+09,fra+11,fro+12}, who used the Bayesian Information Criterion (BIC) to distinguish between cases with and without differential rotation while allowing for spot evolution. This approach is more sophisticated than most, but unfortunately none of the teams involved in the present exercise used it.

\subsection{Limitations of this blind test}

The light curve simulations used in this exercise had a number of limitations, which one might want to address in any future exercise with similar goals.

First, the distribution of simulated periods and amplitudes were chosen to match those found by recent studies based on \emph{Kepler} data. This made sense at the time, but as a result the conclusions of the present exercise apply only within the range of parameter space probed, in particular, periods ranging from 1 to 50 days and activity levels from 0.3 to 3 times solar.  In hindsight, it would have been interesting to go to longer periods and smaller amplitude, to enable us to test whether the observed distributions are real, or the result of selection effects. A significant number of light curves in the \emph{Kepler} data show long-lived spot groups that live for many rotations of the star. We do not model such persistent spot groups in our simulated light curves; rather, our investigation was more focused on spot groups with half-lives up-to ten stellar rotations. Simulating light curves of longer lived spots would be interesting since this would negate the confusion between spot evolution and the detection of differential rotation. 

\begin{figure}
  \centering
\includegraphics[width=\linewidth]{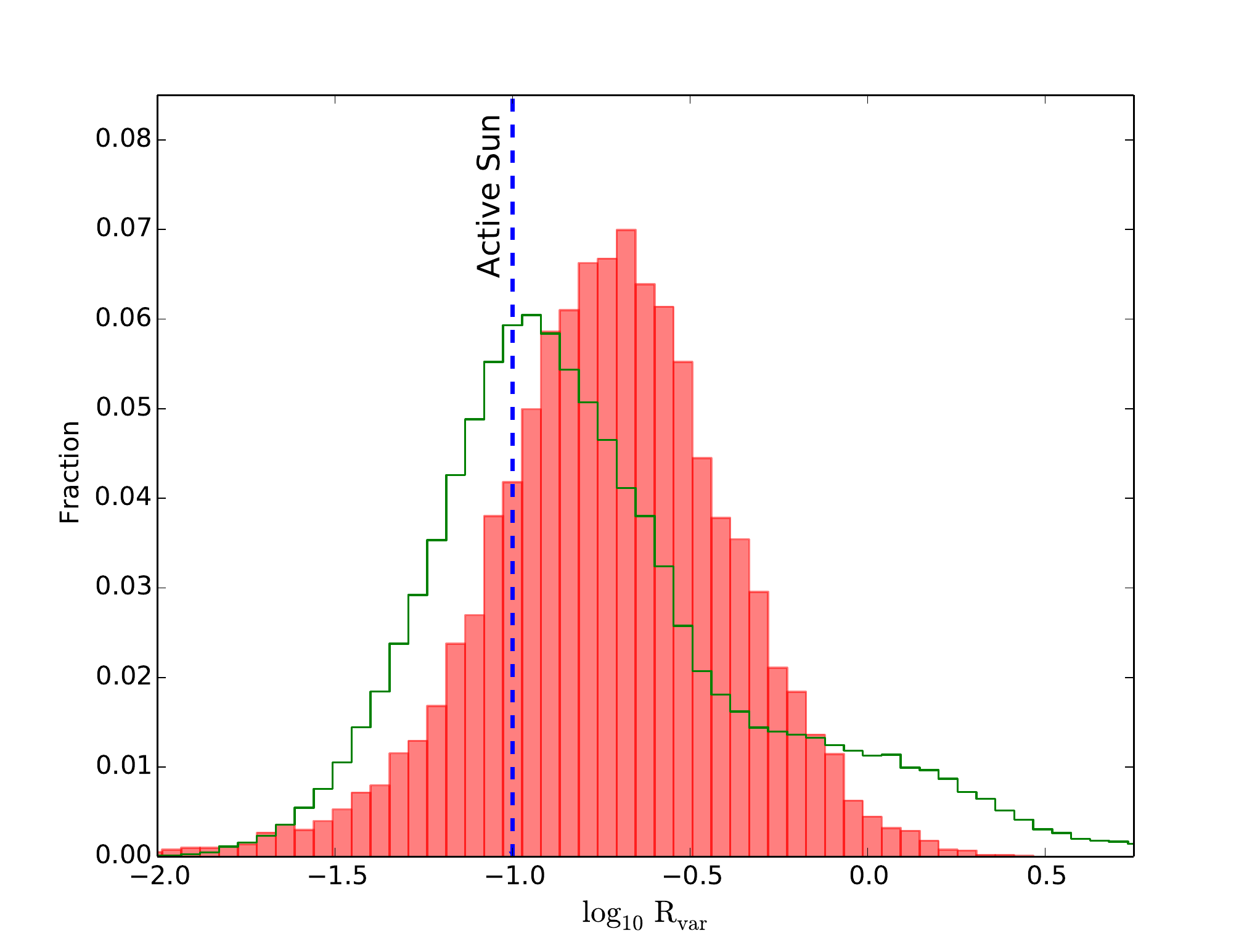}
  \caption{Distribution of the variability in our simulated light curves (red). Overplotted in green is the corresponding distribution for all \emph{Kepler} Q3 stars from \citet{rei+13}. Note that both histograms are normalised such that the total area equals one. The dashed-blue line shows the variability level of the Sun at solar maximum for reference.}
  \label{fig:hist_activity}
\end{figure}

Furthermore, the sample did not include any very active stars (the activity rate was limited to 3 times solar or lower). Because the activity rate controls the number of spots emerging per unit time, while the spot lifetimes were specified as a multiple of the period, this results in particularly small amplitudes for the short-period cases (as can be seen in the right column of Figure~\ref{fig:no_teams}). Thus, the most promising type of stars for differential rotation studies, namely stars with large-amplitude, short period rotational modulation, were absent from our sample. This makes it difficult to compare our results to those of previous studies attempting to measure differential rotation, which have focused mainly on such stars. 

Figure \ref{fig:hist_activity} shows the variability in our simulated light curves derived by splitting each light curve into 90 day windows. We then fit a low order polynomial to each time window and calculate the range. Over plotted is the analogous distribution from all \emph{Kepler} Q3 stars from \citet{rei+13}. The histograms show how our sample does not cover the most active stars in the \emph{Kepler} survey. The results presented here should therefore only be thought of in the context of stars exhibiting low and moderate levels of activity. 

\section{Acknowledgments}
The authors would like to thank the referee, Prof. G. Basri, for suggesting several ways to improve our discussion and asking thought-provoking questions that helped us to improve the final manuscript. The authors are grateful to All Souls College Oxford, for supporting and hosting a 3-day workshop during which the present paper was finalised. SA's contribution to this work was supported by the UK Science and Technology Facilities Council through Consolidated Grant ST/K00106X/1. SA also wishes to thank I.\ Ribas for pointing out a couple of small errors, which had crept into the expression for $f$ \citet{aig+12} (these errors were corrected in the present paper). JL acknowledges support through NASA/GALEX grant program under Cooperative Agreement No. NNX12AC19G issued through the Office of Space Science. MLC acknowledges a CAPES/PNPD fellowship. JRM and M.L.C. acknowledge financial support of the INCT INEspa\c{c}o. TC and RAG wants to acknowledge the funding of the CNES grant at the CEA, as well as the ANR (Agence Nationale de la Recherche, France) program IDEE (n ANR-12-BS05-0008) ``Interaction Des Etoiles et des Exoplanetes''.
 
\bibliography{diffrot}
\bibliographystyle{mn2e}

\bsp

\label{lastpage}

\onecolumn
Full version of Table 2 to be available online only. Notes: ($^a$) $A$ is relative to solar; ($^b$) If $R$ is 0, spots follow a butterfly pattern, otherwise they are randomly distributed in latitude; ($^c$) A KID of 0 means that the noise-free light curve was used, while the entry `Sun' in this column means that the light curve is based on a segment of the Sun's observed total irradiance variations (see Section~2.4 for details).
\begin{longtable}{rrrrrrrrrrrrrrr}
 \hline
$N$ & $A$ & $C_{\rm len}$ & $C_{\rm over}$ & $\theta_{\rm min}$ & $\theta_{\rm max}$ & $R$ & $\sin i$ & $P_{\rm eq}$ & $\delta\Omega/\Omega_{\rm eq}$ & $\tau_{\rm dec}$ & KID & $P_{\rm obs}$ & $P_{\rm obs,min}$ & $P_{\rm obs,max}$ \\ & ($^a$) & (yr) & (yr) & ($^{\circ}$) & ($^{\circ}$) & ($^b$) & & (day) & & ($P_{\rm eq}$) & ($^c$) & (day) & (day) & (day) \\
\hline \hline
1&  0.39&  1.12&  0.27&  12.94&  75.53&  0&  0.19&  1.1&  0.46&  8.20&  893468&  1.51&  1.26&  1.77  \\
2&  0.45&  9.52&  0.19&  19.81&  62.64&  0&  0.99&  1.58&  0.40&  7.31&  1163450&  1.87&  1.75&  2.02  \\
3&  1.34&  1.54&  1.53&  32.99&  74.14&  0&  0.34&  17.04&  0.00&  1.02&  1431090&  17.04&  17.04&  17.04  \\
4&  0.52&  1.09&  0.95&  3.32&  31.71&  0&  0.82&  42.29&  0.61&  3.66&  1849570&  45.44&  43.27&  48.07  \\
5&  0.61&  3.45&  0.12&  27.36&  77.07&  1&  0.55&  44.94&  0.55&  2.39&  2016950&  70.03&  53.33&  89.21  \\
6&  0.73&  1.58&  0.71&  39.96&  66.28&  1&  0.81&  23.64&  0.00&  2.13&  2017080&  23.64&  23.64&  23.64  \\
\end{longtable}


\end{document}